\documentclass[useAMS,usenatbib]{mnras}
\usepackage{amssymb}
\usepackage{natbib}
\usepackage{times}
\usepackage{graphics,epsfig}
\usepackage{graphicx}
\usepackage{amsmath}
\usepackage{amssymb}
\usepackage{comment}
\usepackage{supertabular}

\usepackage{graphicx}
\usepackage{amsmath}
\usepackage{color}
\usepackage{subfig}
\usepackage{float}

\newcommand{\lya}  {\ensuremath{{\rm Ly}\alpha}}

\newcommand{\kms}{km~s$^{-1}$}

\newcommand{\acc}{\ensuremath{\rm atoms~cm^{-2}}}

\newcommand{\HI}{H{\sc i}}
\newcommand{\tb}{\ensuremath{\rm T_{B}}}
\newcommand{\ts}{\ensuremath{\rm T_{s}}}
\newcommand{\tk}{\ensuremath{\rm T_{k}}}

\title[The temperature of the neutral ISM in the Galaxy]{The temperature of the neutral Interstellar Medium in the Galaxy}
\author [Patra et al.]{	Narendra Nath Patra$^{1}$ \thanks {E-mail: naren@iiti.ac.in} and Nirupam Roy$^2$\\
	$^{1}$ Department of Astronomy, Astrophysics and Space Engineering, Indian Institute of Technology Indore, 453552, India \\ $^{2}$ Department of Physics, Indian Institute of Science, Bengaluru 560012, India \\
}
\date {}
\begin {document}
\maketitle

\begin{abstract}

Atomic Hydrogen-21 cm transition (\HI) is an excellent tracer to study and understand the properties of the atomic gas in the Galaxy. Using the Westerbork Synthesis Radio Telescope (WSRT), we observed 12 quasar sightlines to detect galactic \HI~in absorption. We achieve an optical depth RMS of $\sim 1-2 \times 10^{-3}$, essential to detect the Warm Neutral Medium (WNM). We detect \HI~absorption in all our sightlines except along 1006+349, for which we set a strict upper limit on the spin temperature as $\langle T_s \rangle > 570$ K. We find around 50\% of our sightlines have $\langle T_s \rangle > 500$ K, indicating a WNM dominance. Further, we calculate an upper limit of the CNM fraction along our sightlines and find a median CNM fraction of $\sim 0.12$. With our observations, we reconfirm the existence of a threshold column density of $\sim 2 \times 10^{20}$ \acc~to form CNM in the ISM. Using a two-temperature model of the \HI~disk, we explore the distribution of spin temperature in the Galactic ISM. We find that a simple fixed axisymmetric two-temperature model could not produce either the observed column density or the integral optical depth. This indicates the existence of a more complex distribution of spin temperatures in the Galaxy.

\end{abstract}


\section{Introduction}

The neutral component of the Interstellar Medium (ISM) plays a crucial role in galaxy formation and evolution. It acts as a long-term fuel reservoir for star formation and, hence, can significantly influence the physical and chemical properties of the ISM \citep{krumholz2011,glover2012}. Not only that, it almost makes up for half of the ISM in the Galaxy \citep[see, e.g., ][for a review]{mcclure-griffiths2023}. In that sense, it is essential to investigate the conditions and temperatures of different phases of the neutral ISM to understand the physical connection between gas and star formation.

The neutral hydrogen (\HI) largely dominates the neutral part of the ISM, making almost three-fourths of it \citep[see, e.g.,][]{carilli13}. The \HI~in the ISM of galaxies can be found in a range of temperatures \citep{heiles03b}. Theoretical calculations show that in thermal equilibrium, the \HI~would settle in two distinct phases, namely, the Cold Neutral Medium (CNM) and the Warm Neutral Medium (WNM). Gas in these two phases would also be in pressure balance to maintain a steady state. The CNM is the cold component of the ISM with a kinetic temperature of $\lesssim$ 300 Kelvin with a high particle density of $\sim 10-100~cm^{-3}$ \citep{kulkarni88,dickey90}. A stable WNM, on the other hand, is expected to have a kinetic temperature of $\sim 5000 - 8000$ Kelvin with a much lower particle density of $\sim 0.1-1~cm^{-3}$ \citep[see, e.g.,][]{dickey78,payne83,heiles03a,roy06}. Theoretically, the high sensitivity of the heating and cooling processes in the ISM on the ambient temperature leads the thermal runaway processes, such as any gas in the intermediate temperature (i.e., nither in CNM nor WNM phase) quickly moves into one of the stable phases \citep[see, e.g.,][]{field69b,wolfire95b,wolfire03,bialy2019}. In that sense, a minimal amount of gas is expected to be observed in the unstable phase in galaxies. However, recent sensitive observations of the ISM in the Galaxy provide enough indication that a considerable amount of gas might be present in the intermediate unstable phase \citep{heiles03a,heiles03b,kanekar03a,roy13a,murray15,murray2018}. This calls for an understanding of the mechanisms required to drive a significant amount of stable gas into the unstable intermediate phase. Several numerical studies suggest that the typical supernovae rate in a galaxy can significantly influence the amount of gas in CNM and WNM through shock propagation and inducing turbulence in the medium \citep{koyama02,maclow05,kim2013,kim14,gazol2016}. \citet{audit05} showed that turbulent flow collisions in the ISM could produce up to 30\% intermediate gas. While considering only collisions from the wind-blown superbubble, \citet{ntormousi11} have found that only 8-10\% gas could be pushed to the intermediate phase. These studies steer towards the necessity of more observational inputs to understand the abundance and physical processes dictating the phases in the ISM.

The physical conditions of the \HI~in different phases of the ISM can be described by two temperatures, i.e., the excitation temperature or the spin temperature (\ts) and the kinetic temperature (\tk). \ts~determines the level populations of the two hyperfine levels of the \HI~in ground state, whereas the \tk~represents the kinetic temperature of the gas. Determination of these temperatures is crucial to understand the physical processes in the ISM. Though the emission studies in galaxies \citep{young96,younglo97,warren12,patra16,saikia2020,hunter2021,biswas2022} provide spatially resolved details of ISM phases and decidedly confirmed the existence of a two-phase medium through multiple Gaussian decompositions of \HI~emission spectra, these studies can not determine the spin temperature of the gas independently.
 
Nonetheless, a combination of both the absorption and emission studies allows one to determine the spin and the kinetic temperature at the same time \citep[see, e.g.,][]{carilli98,dwarakanath02,kanekar03b,roy13a,roy13b,murray14,murray15,murray2018,patra18b,rybarczyk2020,murray2021,allison2022,hu2023,smith2023}. However, as the absorption studies cut through a pencil beam, it only probes the characteristics of the intervening gas along that sightline. Also, due to the optical depth dominance of the CNM, these studies mostly returned the properties of the CNM. For example, the kinetic temperature of the CNM gas was found to be  $\lesssim 300$ K. Collisional processes and the resonant scattering of the \lya~photons force the \ts~to approach \tk in the dense ISM of the CNM \citep{field58,deguchi85,liszt01}. The WNM (or the intermediate gas), on the other hand, has a low particle density with a high kinetic temperature. This gas consecutively produces very low peak optical depth, which is very often missed in the previous shallow observations.

To examine the properties of the WNM and the intermediate gas in the Galaxy, we have undertaken a large project to detect Galactic \HI~in absorption against background quasars through sensitive interferometric observations. As part of this effort, we reported observations of 32 sightlines previously \citep{roy13a,roy13b,patra18b} with the Giant Meter wave Radio Telescope (GMRT) \citep{swarup91} and the Westerbork Synthesis Radio Telescope (WSRT) \citep{baars1973}. In continuation of the same effort, we observed another 12 sightlines using the WSRT and detected Galactic \HI~in absorption in 11 of them. In this paper, we present our observations, data analysis, and the resulting spectra. We also use the emission spectra from the Leiden-Argentine-Bonn (LAB) survey \citet{kalberla08} along our sightlines to further investigate the properties of the ISM in the Galaxy.

\section{Observation and data analysis}

\begin{figure}
\begin{center}
\begin{tabular}{c}
\resizebox{0.45\textwidth}{!}{\includegraphics{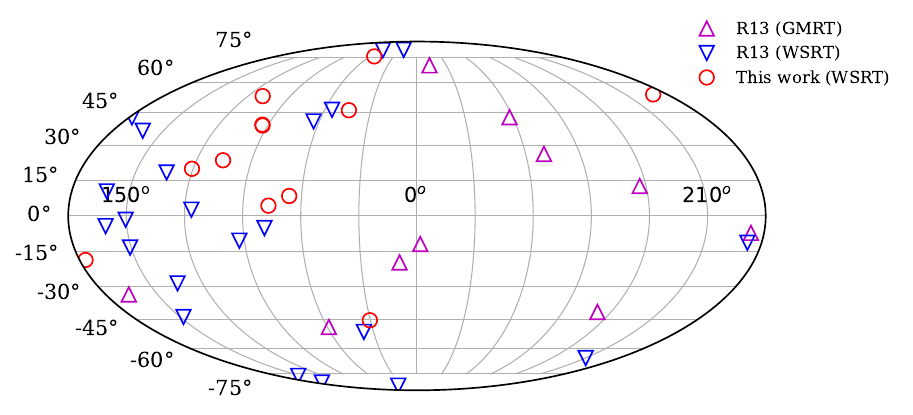}}
\end{tabular}
\end{center}
\caption{Distribution of the target line-of-sights for detecting galactic \HI~absorption is shown in the Mollweide projection of Galactic coordinates. R13 in the legend represents \citet{roy13a}. As can be seen, the sightlines are randomly distributed in Galactic latitude, eliminating any bias on the statistical properties of the ISM. See the text for more details.}
\label{skyplot}
\end{figure}

\begin{figure*}
\begin{center}
\begin{tabular}{c|c|c}
\resizebox{.33\textwidth}{!}{\includegraphics{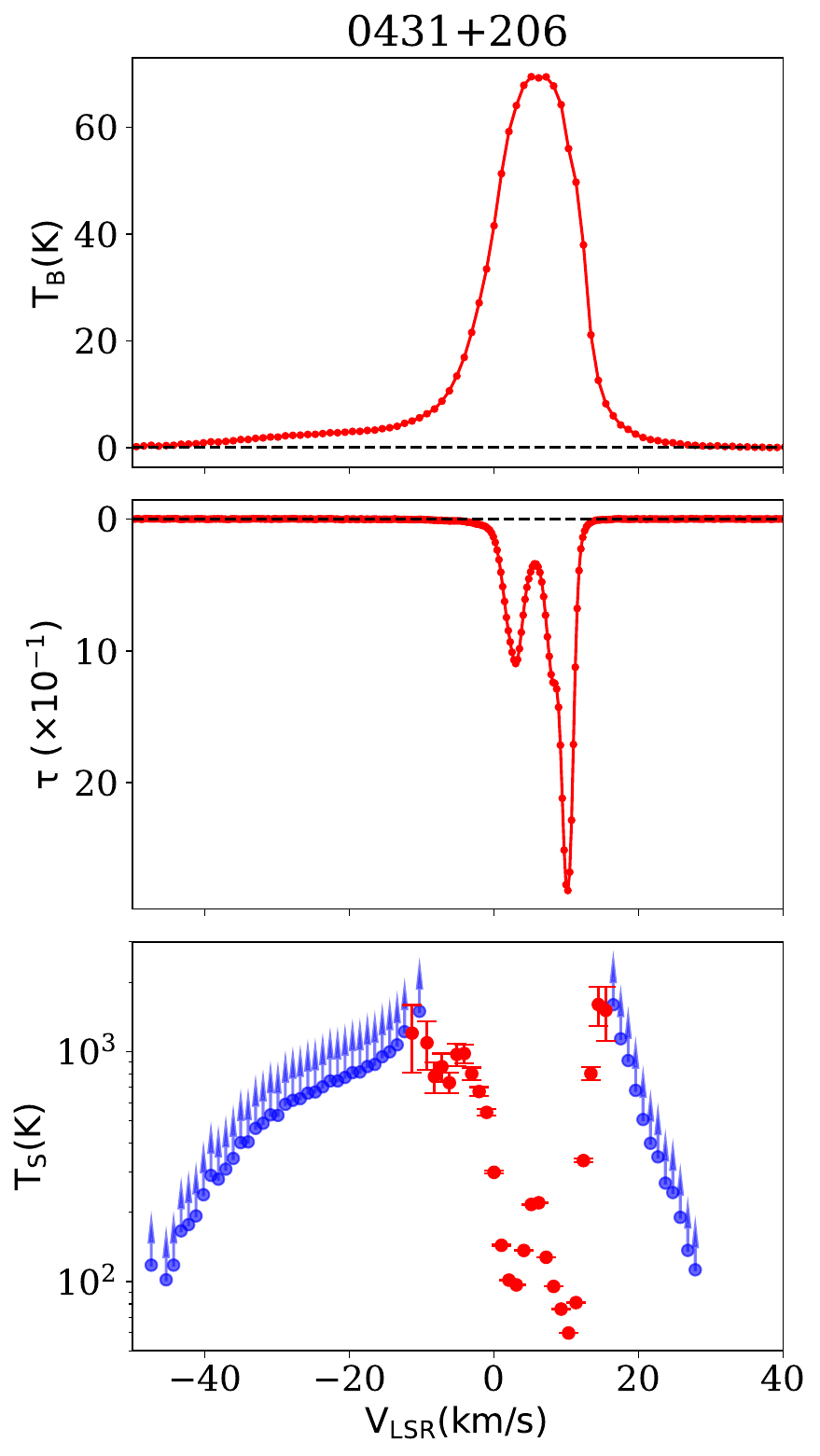}} &
\resizebox{.33\textwidth}{!}{\includegraphics{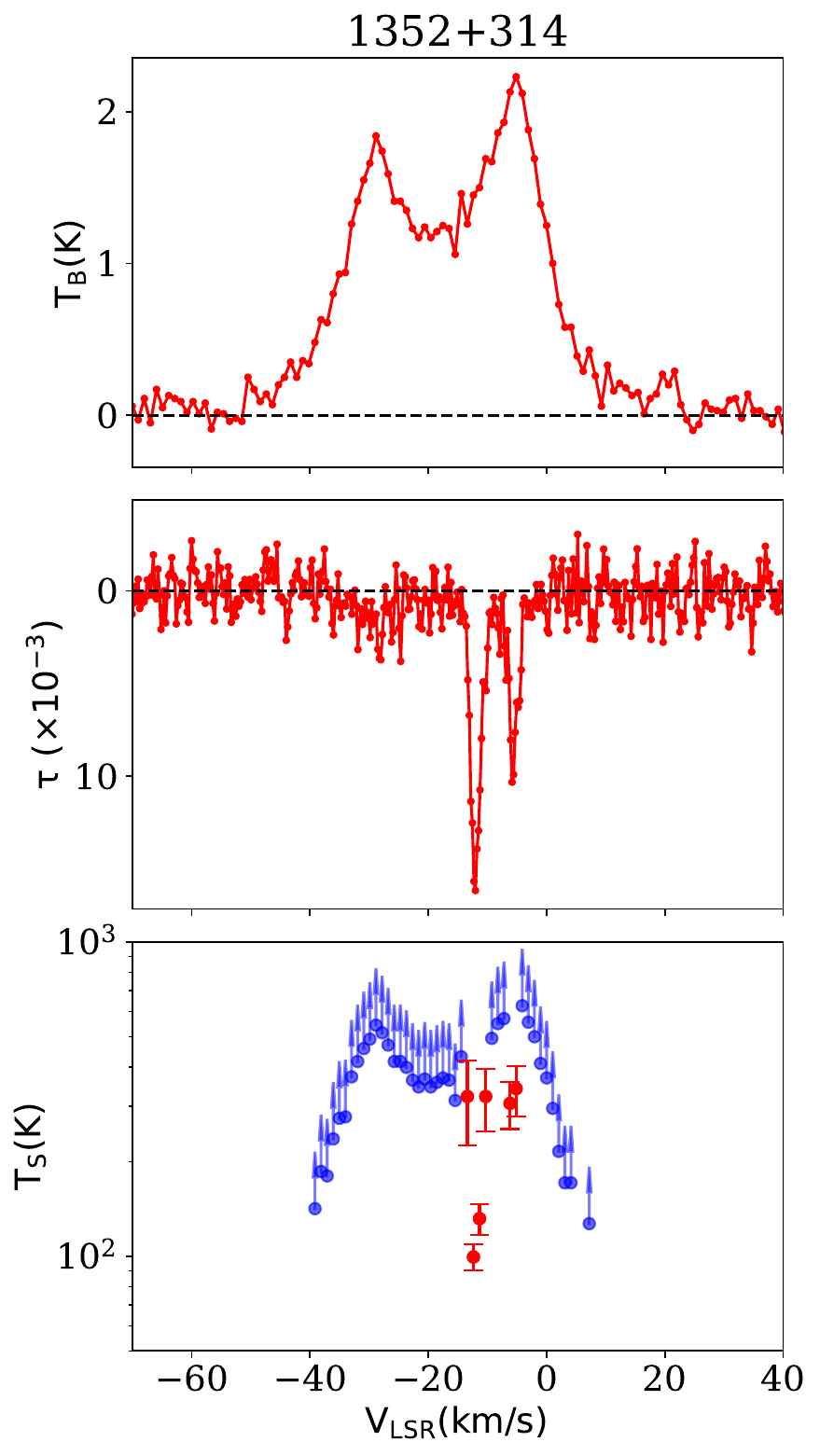}} &
\resizebox{.32\textwidth}{!}{\includegraphics{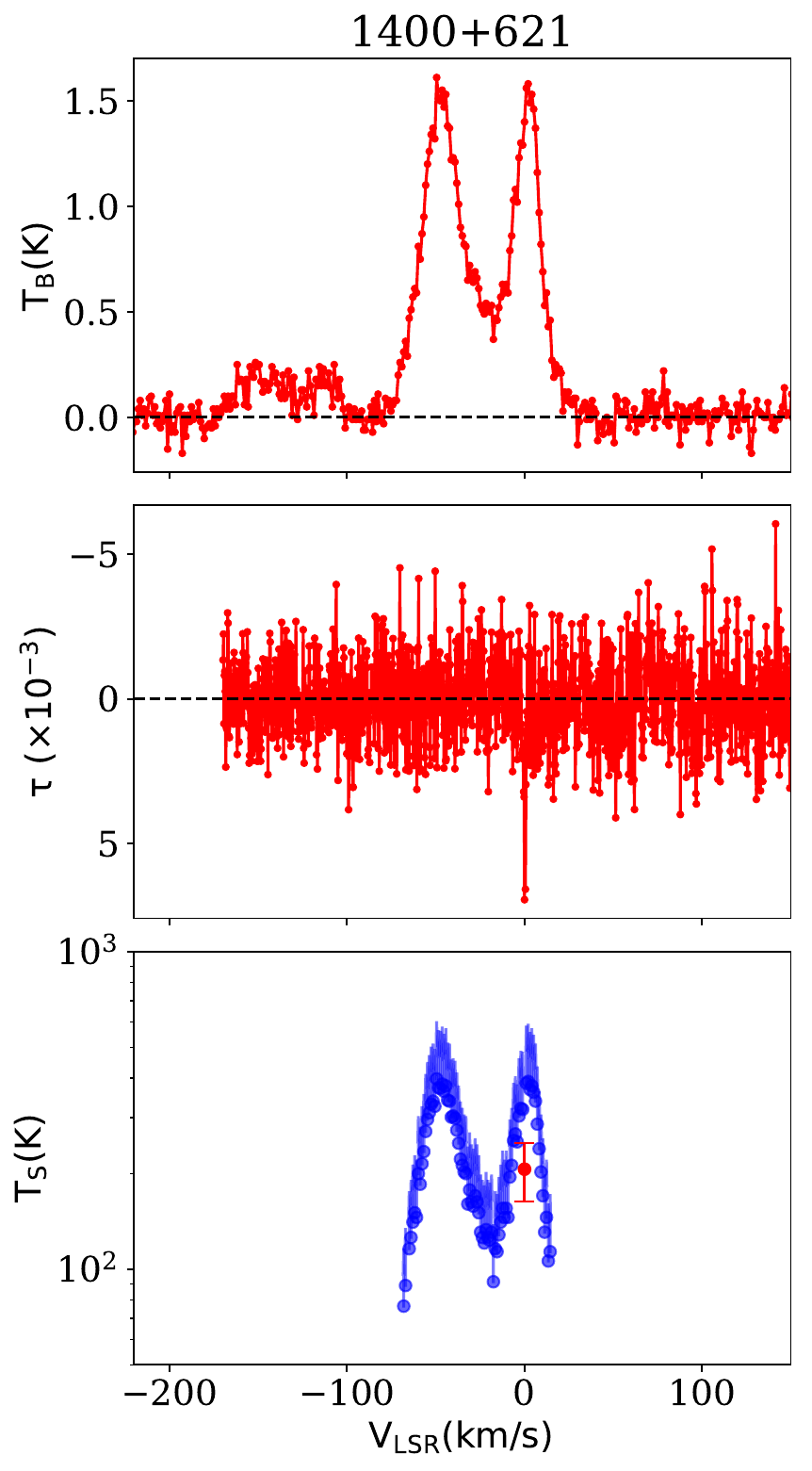}} \\

\resizebox{.35\textwidth}{!}{\includegraphics{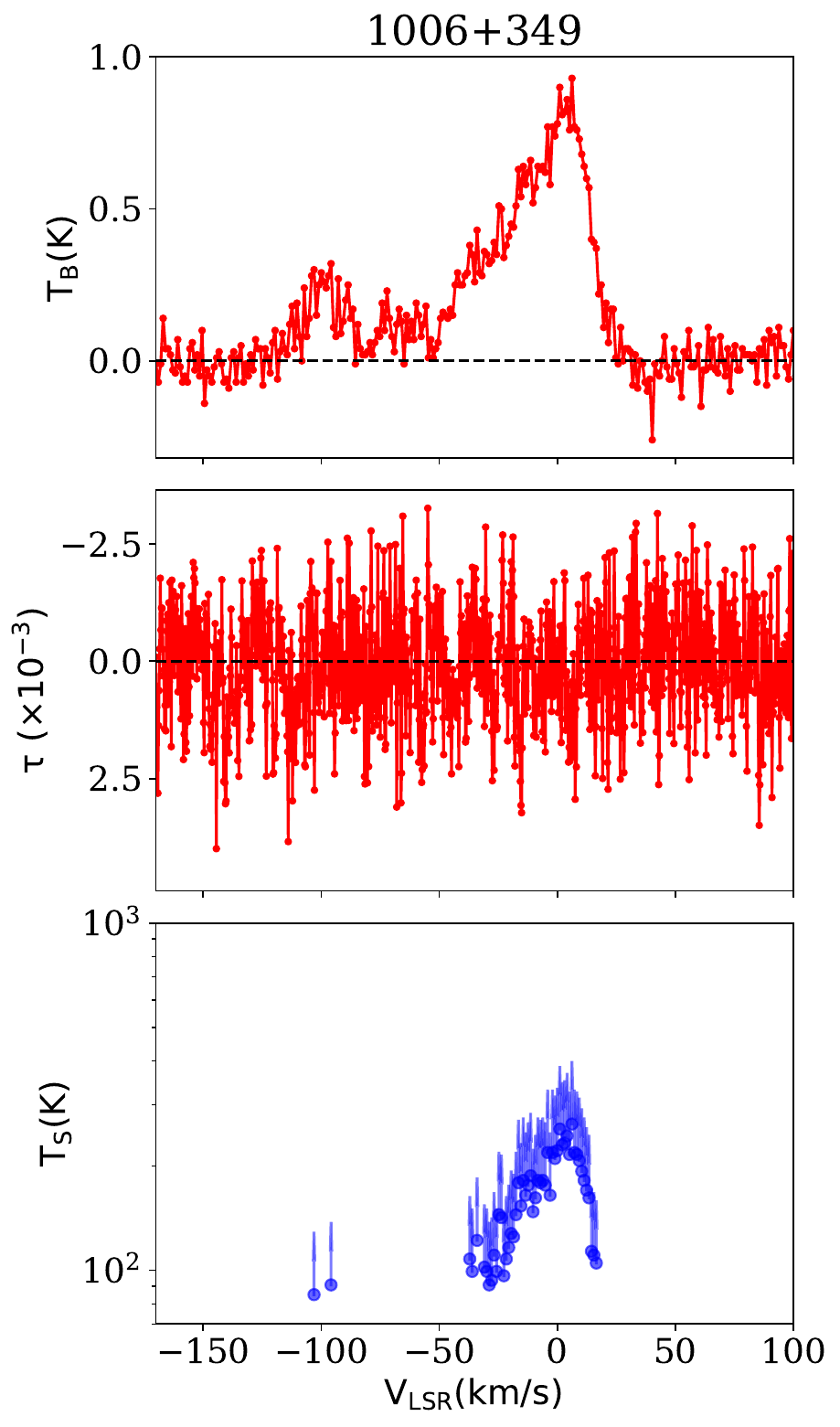}} &
\resizebox{.33\textwidth}{!}{\includegraphics{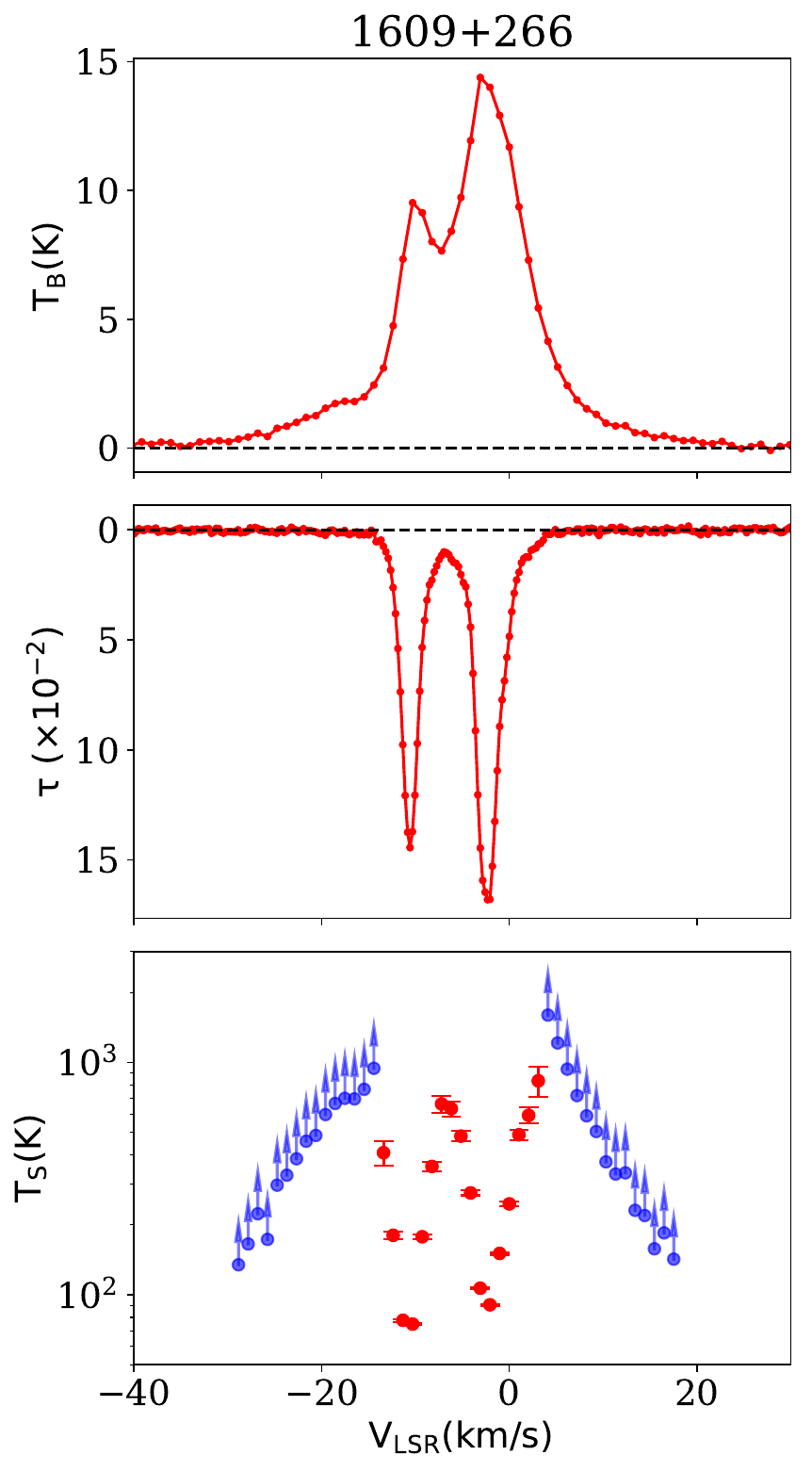}} &
\resizebox{.33\textwidth}{!}{\includegraphics{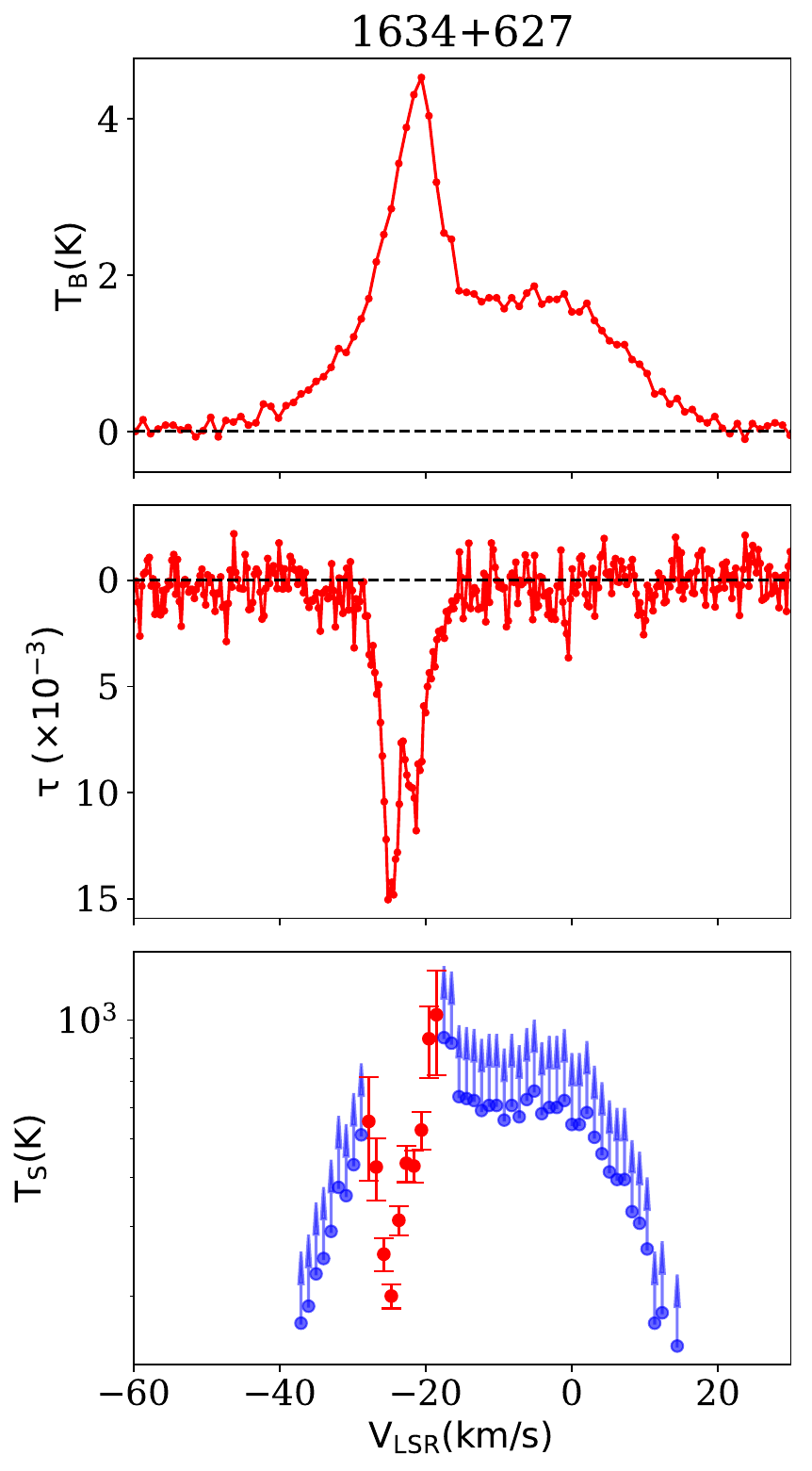}} \\
\end{tabular}
\end{center}
\caption{The spin temperature spectra of our target sight-lines. The top panels show the brightness temperature spectra observed in \HI~emission from the LAB survey. The middle panels represent the optical depth spectra as obtained from our observations. The bottom panels show the spin temperature spectra. The red solid circles with error bars represent spin temperatures where both the optical depth and the brightness temperature could be measured with more than 3-sigma significance. Whereas, the blue circles with the up arrows indicates the lower limits of the spin temperature. At these velocities, only \tb~could be detected at more than 3-sigma significance but the $\tau$ is not detected with more than 3-sigma significance.}
\end{figure*}

\begin{figure*}
\ContinuedFloat
\begin{center}
\begin{tabular}{c|c|c}
\resizebox{.33\textwidth}{!}{\includegraphics{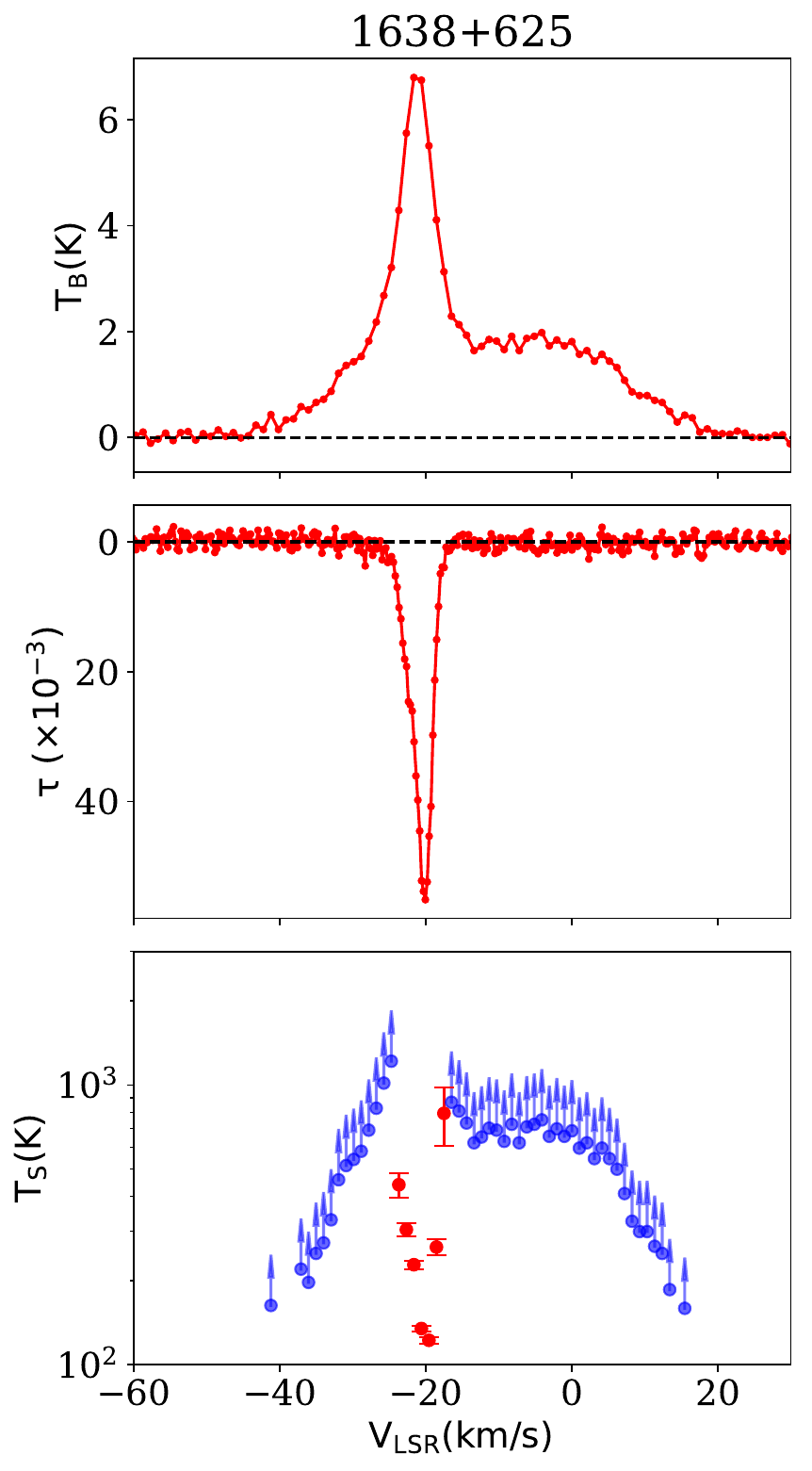}} &
\resizebox{.33\textwidth}{!}{\includegraphics{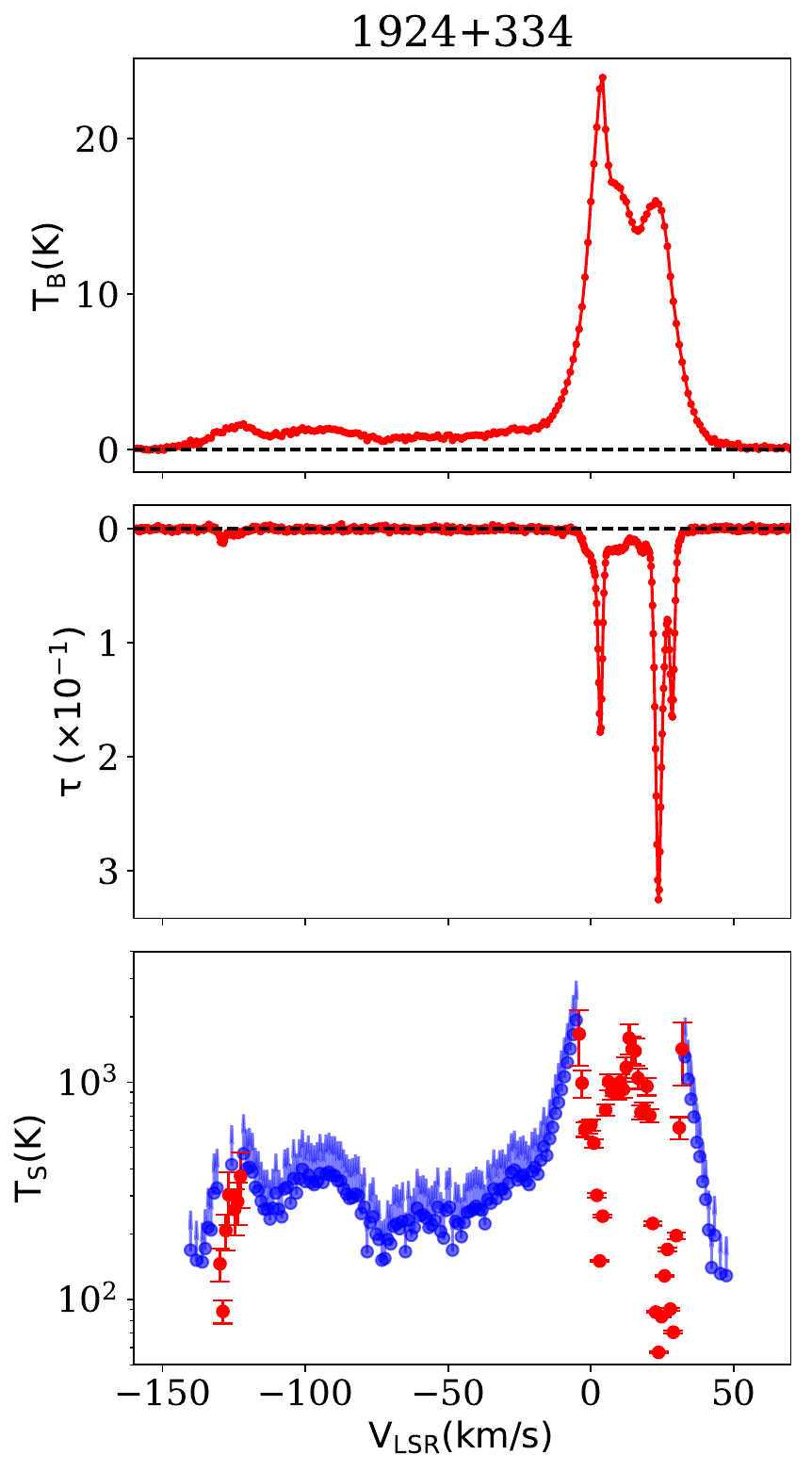}} &
\resizebox{.33\textwidth}{!}{\includegraphics{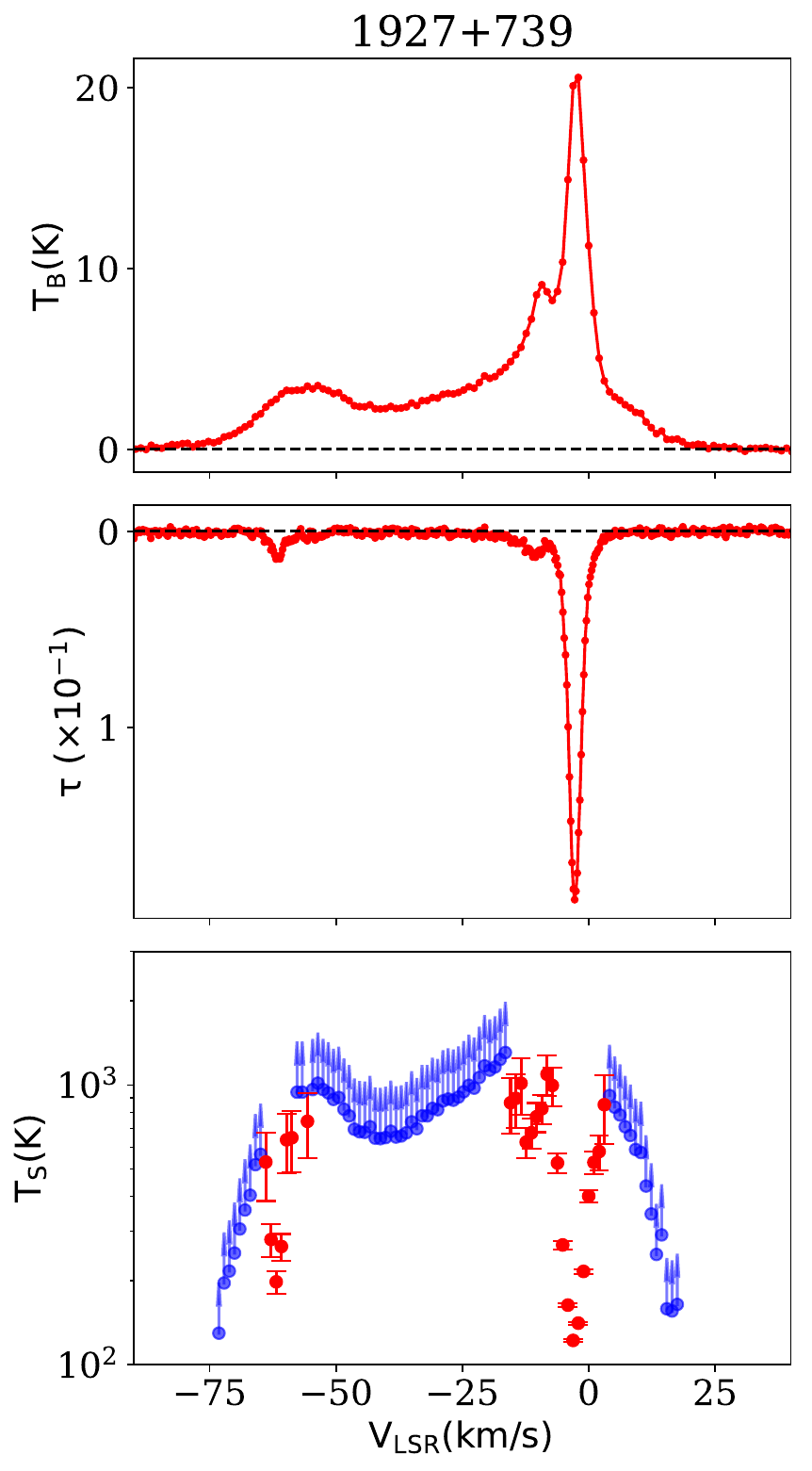}} \\

\resizebox{.33\textwidth}{!}{\includegraphics{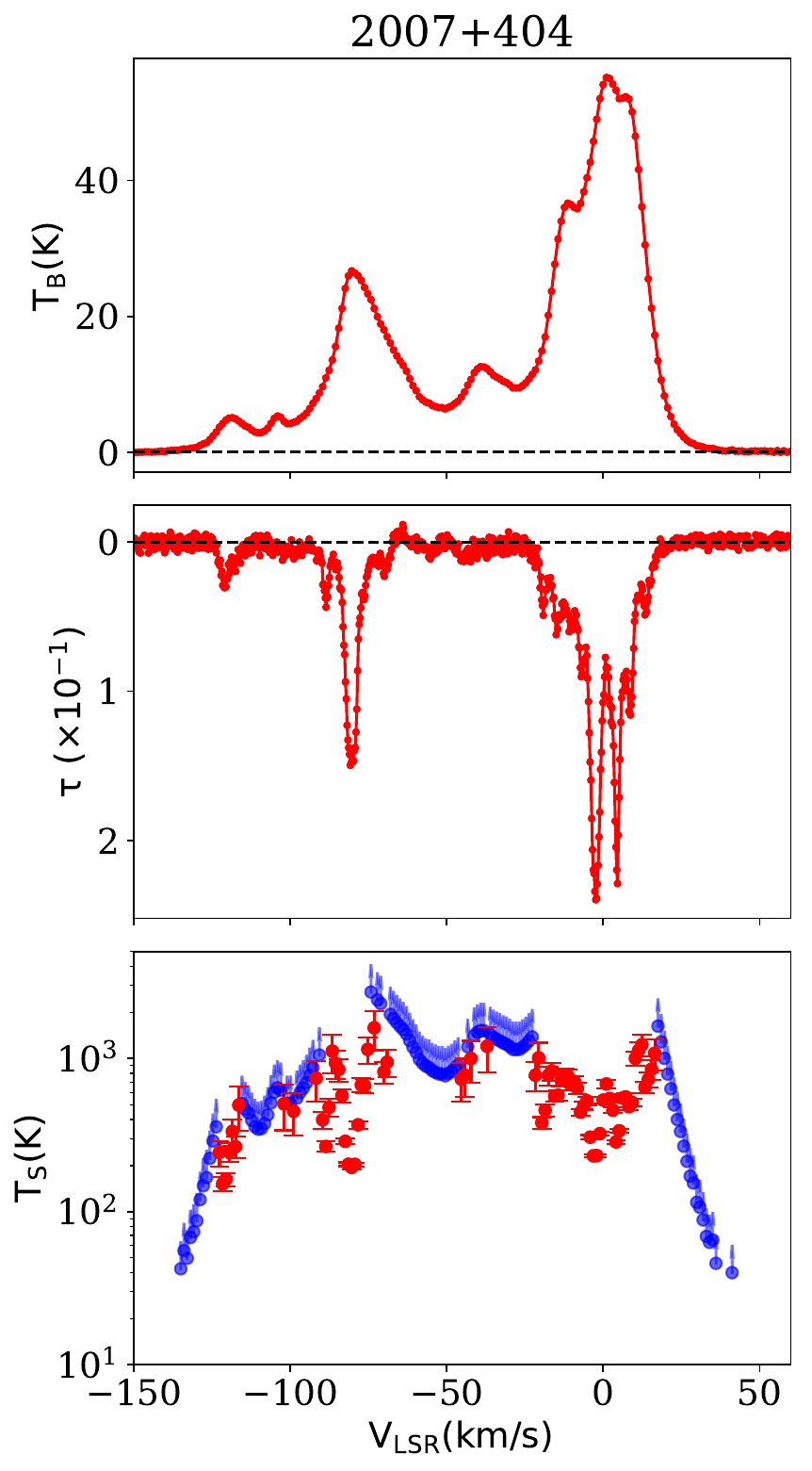}} &
\resizebox{.33\textwidth}{!}{\includegraphics{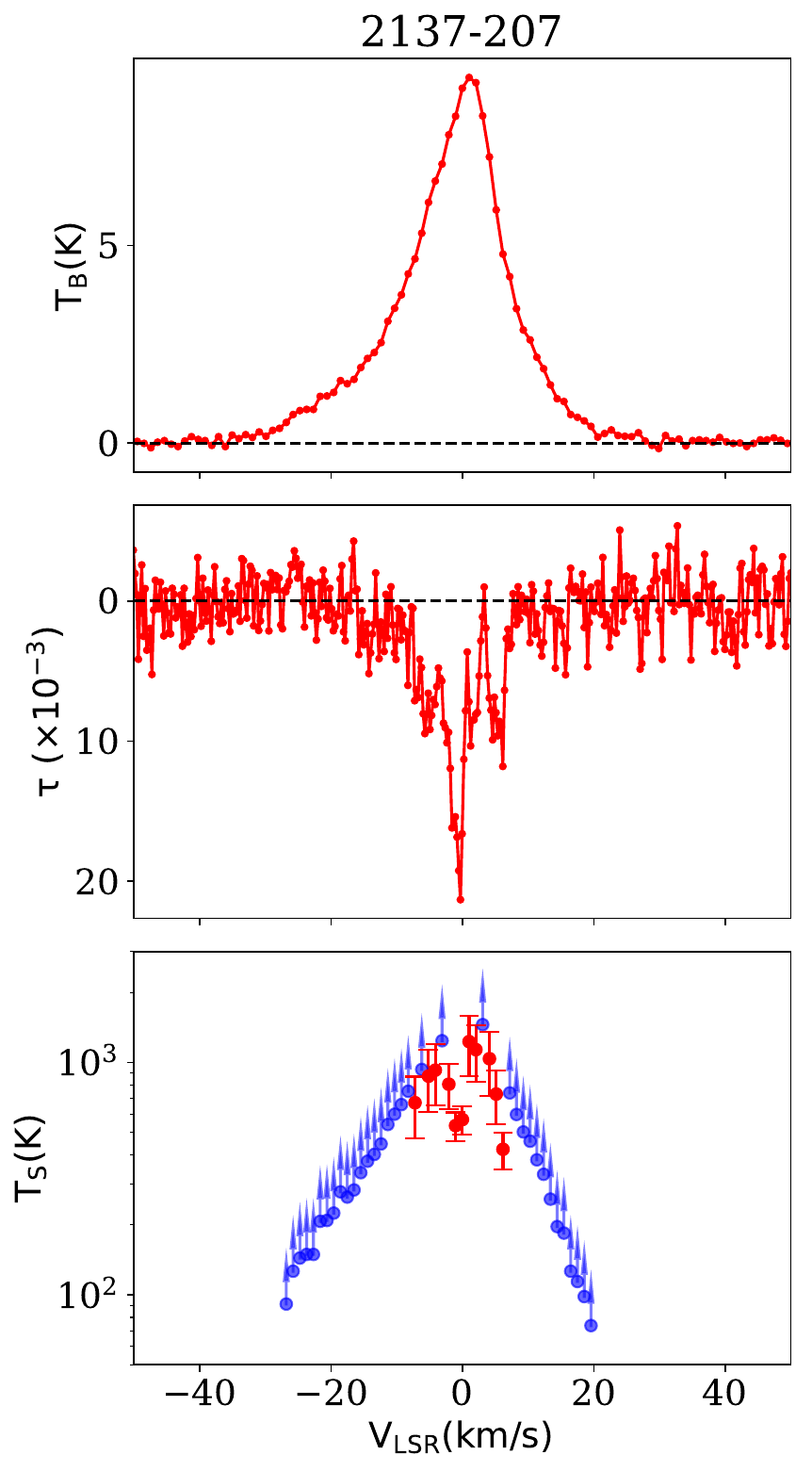}} &
\resizebox{.33\textwidth}{!}{\includegraphics{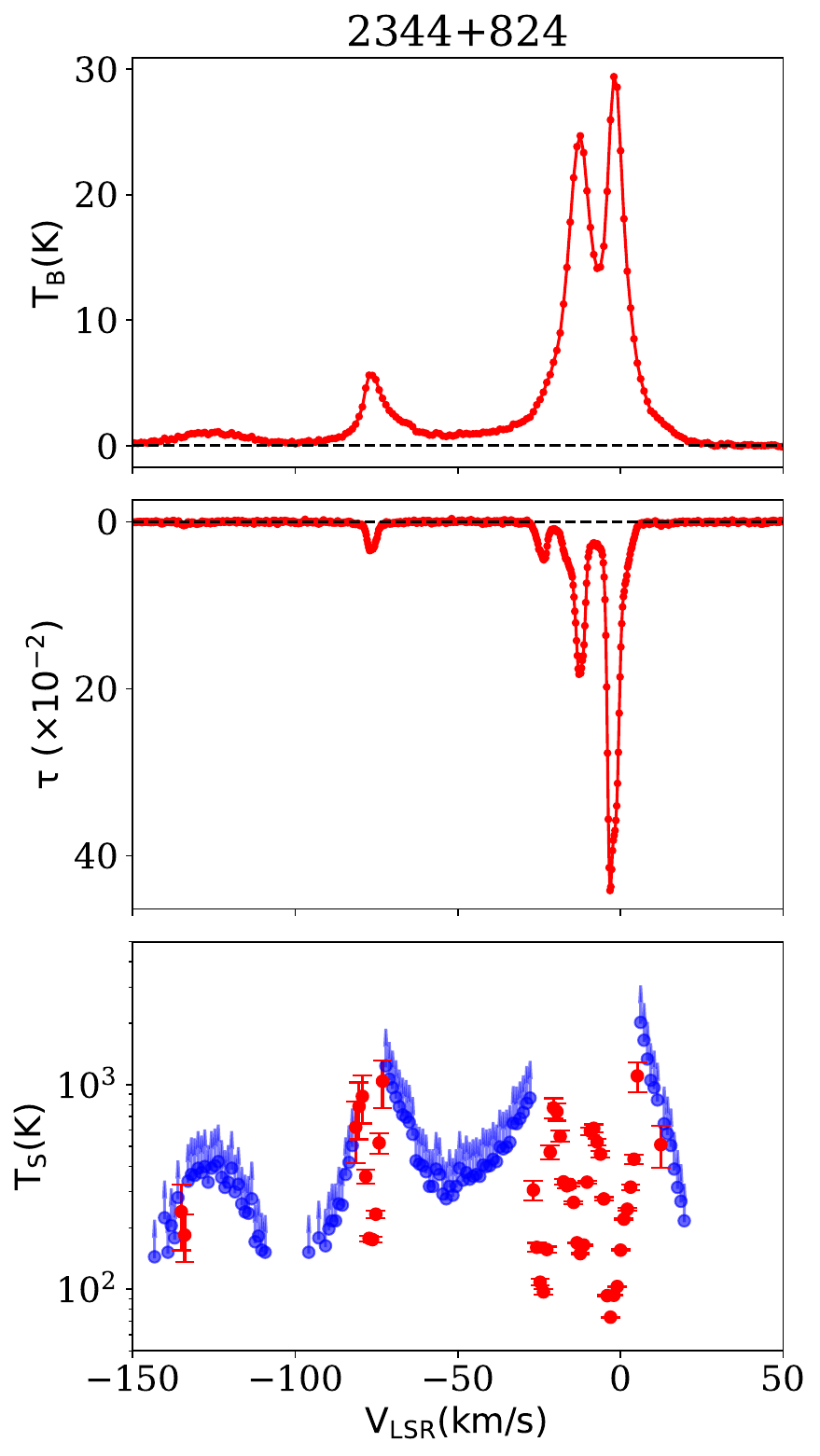}} \\

\end{tabular}
\end{center}
\caption{Continued.. Fig.2}
\label{fig:ts}
\end{figure*}

In Fig.~\ref{skyplot}, we show the distribution of our new (empty red circles) and old (empty up and down triangles) target sightlines in the sky. As can be seen from the figure, our target directions are randomly distributed in Galactic latitude, which eliminates any bias on the statistical properties of the ISM. All our sightlines are selected, such as the background souces to have a flux $\gtrsim 3$ Jy at 1.4 GHz. A high flux of the background continuum source results in a lower optical depth RMS. However, due to the paucity of the high flux density backgrounded sources on the sky, one must resort to more extended integration to achieve a low optical depth RMS. For typical ISM conditions, a 5-sigma detection of the WNM requires an optical depth RMS of $\sim \ 10^{-3}$. Not only that, due to high kinetic temperature, the width of the WNM spectra is also high with low peak optical depth. At the same time, almost all the sightlines host high optical depth CNM along with the WNM or intermediate gas. These ISM conditions require not only low optical depth RMS but also a high dynamic range and high spectral resolution. Our sensitive observations accomplish all these conditions making it feasible to detect the WNM or intermediate gas in the Galaxy. 

We use the Westerbork Synthesis Radio Telescope (WSRT) to observe a sample of 12 sightlines to detect Galactic \HI~in absorption. The observations were carried out between June and November of 2008. All the sources were observed with a bandwidth of 2.5 MHz with 2048 channels or 5 MHz with 4096 channels yielding a spectral resolution of $\sim 0.26$ \kms. An in-band frequency-switching technique is used for bandpass calibration. In this technique, the frequency of the observation was switched every 5 minutes, and the data were recorded in separate IFs. The switching frequency was calculated carefully, such as the line always remains inside the observing band. This provides a unique advantage over the position switching as in position switching, half of the time is lost during off-source observation. However, it should be mentioned here that, for frequency-switching to work, one must have a flat bandpass response of the telescope. Otherwise, any residual structure over flat bandpass will introduce an artificial shape (residual band shape) in the line over applying the bandpass calibration. We find the band shapes of all the WSRT antennas are flat such as it can be used to perform frequency-switching observations. All our sources were observed with an on-source time of 12 hours resulting in an optical depth RMS sensitivity of $\sim 10^{-3}$ per 0.26 \kms~channel, which is essential for detection of high \ts~gas in an absorption spectrum.

All the data are analyzed using the tasks of the standard Astronomical Image Processing System (AIPS) \citep{vanmoorsel1996,jacoby1996} and the Common Astronomy Software Application (CASA) \citep{casa2022}. Firstly, the visibilities are inspected and edited to remove any bad data due to Radio Frequency Interference, system failure, etc. After the editing of the data, a flux calibration is performed and applied to the data using the standard flux calibrators on the WSRT sky (3C48 or 3C147 or 3C286). The flux calibrators were observed for 10 minutes in the beginning and at the end of every observing run. For our observations, the target source itself is used as a phase and bandpass calibrator. It should be mentioned here that both the IFs are calibrated separately for flux, phase, and bandpass calibration. However, in each IF, some of the channels would be affected by the presence of the absorption line. In these channels (for a particular IF), the bandpass solutions would not be accurate. However, due to frequency switching, the channel numbers affected by the absorption line would be different in different IFs. Assuming a flat bandpass, we then exchange the bandpass tables and apply it to the IFs. Thus, the correct bandpass solutions are applied to the channels containing the absorption line in each IF. After all the calibrations applied, the continuum is subtracted from the visibility data by fitting a linear curve to the visibilities of line-free channels, using the AIPS task {\tt UVLIN}. 

Thus obtained continuum subtracted visibilities are then imaged to extract the spectrum along the quasar sightline. However, due to the requirement of high spectral resolution, the visibility cubes contain a large number of spectral channels (2048 or 4096). Imaging this large number of spectral channels is compute-intensive and hard to perform in AIPS as it does not implement parallel processing. To avoid the same, we perform the imaging in CASA, which works faster as it can image different spectral channels in parallel. The Doppler correction to the data was performed on the fly during imaging. The imaging was done using a natural weighting, and no cleaning was done as we are interested in the point source at the phase center. The frequency axes of the image cubes are then converted to velocities with respect to the local standard of rest ($V_{LSR}$), and a spectral cut is taken along the position of the compact point source at the phase center. Thus produced absorption spectra are then converted into optical depth spectra using the flux density of the background source as determined from the continuum maps. In Fig.~\ref{fig:ts} middle panel, we show the optical depth spectra of our target sightlines.

We calculate the noise on the absorption spectra, adopting a similar approach, as described in \citet{roy13a}. The total noise on an absorption spectrum has a contribution from two components. The first component ($n_c$) originated due to the system temperature, which depends on the frequency-dependent sky brightness temperature. The second component ($n_b$) originates due to the errors in bandpass (could be due to poor band stability, error in calibration, etc.). The sky brightness temperature (and hence the $n_c$) would be higher at frequencies where there is emission from galactic \HI. We assume that, at frequencies, where there is no emission or absorption, these two components will contribute equally in quadrature to the noise. Hence, we estimate the noise in line-free channels and calculate $n_c$ and $n_b$. The $n_b$ is frequency independent, and hence, it would be the same over the full spectrum. But, $n_c$ would raise by a factor of $(T_{sky} + T_{sys})/T_{sys}$, due to enhanced sky brightness due to galactic \HI~emission. We obtain the $T_{sky}$ from the LAB emission spectra, and $T_{sys}$ for WSRT at L-band is $\sim$ 30 K. The final noise in any channel is then estimated by adding these two components ($n_c$ and $n_b$) in quadrature.

\section{Results and Discussion}

\subsection{CNM/WNM fraction in the ISM of the Galaxy}

\begin{table*}
\caption[obs_result]{Summary of observation results}
\begin{tabular}{l|c|c|c|c|c|c|c|c|c|c}
\hline
Name & l & b & $ S_{NVSS}$ & $\tau_{rms}$ & $\tau_{peak}$ & $\int \tau dv$ & $ N_{HI}$ & $<T_s>$ & $\Delta V^{em}_{90}$ & $\Delta V^{abs}_{90}$ \\
 & (deg) & (deg) & (Jy) & ($\times 10^{-3}$) & ($\times 10^{-3}$) & ($ kms^{-1}$) & ($\times 10^{20}$) & (K) & ($ kms^{-1}$) & ($ kms^{-1}$) \\
 \hline
0431+206 & $ 176.8$ & $ -18.6$ & $ 3.9$ & $1.13$ & $2816.400$ & $12.281 \pm  0.005$ & $  19.9$ & $89 \pm 0.04$ & $28$ & $9$ \\ 
1006+349 & $ 190.1$ & $  54.0$ & $ 3.3$ & $1.20$ & $<3.6$ & $<0.077$ & $0.8$ & $>570$ & $114$ & $-$ \\ 
1352+314 & $  54.6$ & $  76.1$ & $ 3.5$ & $1.12$ & $16.170$ & $ 0.079 \pm  0.004$ & $   1.2$ & $834 \pm 40.40$ & $43$ & $26$ \\ 
1400+621 & $ 109.6$ & $  53.1$ & $ 4.4$ & $1.40$ & $6.94$ & $0.009 \pm 0.002$ & $1.6$ & $9875 \pm 1980$ & $149$ & $2$ \\ 
1609+266 & $  44.2$ & $  46.2$ & $ 4.8$ & $0.87$ & $168.050$ & $ 0.963 \pm  0.003$ & $   3.8$ & $218 \pm 0.64$ & $35$ & $12$ \\ 
1634+627 & $  93.6$ & $  39.4$ & $ 4.8$ & $0.93$ & $15.040$ & $ 0.090 \pm  0.002$ & $   1.8$ & $1097 \pm 28.25$ & $120$ & $11$ \\ 
1638+625 & $  93.2$ & $  39.0$ & $ 4.5$ & $0.89$ & $55.100$ & $ 0.186 \pm  0.001$ & $   2.3$ & $670 \pm 4.92$ & $144$ & $4$ \\ 
1924+334 & $  66.4$ & $   8.4$ & $ 3.9$ & $1.24$ & $325.450$ & $ 2.137 \pm  0.005$ & $  13.5$ & $347 \pm 0.84$ & $136$ & $27$ \\ 
1927+739 & $ 105.6$ & $  23.5$ & $ 3.1$ & $1.17$ & $187.660$ & $ 0.818 \pm  0.006$ & $   7.0$ & $469 \pm 3.28$ & $72$ & $60$ \\ 
2007+404 & $  76.8$ & $   4.3$ & $ 3.7$ & $2.66$ & $239.650$ & $ 4.473 \pm  0.024$ & $  47.5$ & $582 \pm 3.10$ & $108$ & $101$ \\ 
2137-207 & $  30.3$ & $ -45.6$ & $ 3.6$ & $1.95$ & $21.300$ & $ 0.153 \pm  0.007$ & $   3.0$ & $1091 \pm 48.79$ & $31$ & $24$ \\ 
2344+824 & $ 120.6$ & $  19.9$ & $ 3.8$ & $0.91$ & $441.380$ & $ 3.121 \pm  0.006$ & $  11.9$ & $209 \pm 0.39$ & $95$ & $26$ \\
\hline
\end{tabular}
\label{sample}
\end{table*}

In table~\ref{sample}, we summarise a few essential parameters computed from the absorption spectra of our target sightlines. In column (1), we present the names of the background sources, whereas columns (2) and (3) represent their galactic coordinates in degrees. Column (4) shows the flux of the continuum source in the L-band as obtained from the NVSS catalog. Columns (5) and (6) present the RMS and peak optical depth, respectively. Column (7) quotes the integral optical depth ($\int \tau dv$) over the absorption line, whereas column (8) presents the \HI~emission column density towards the sightlines as obtained from the LAB survey adopting an optically thin approximation. We note that the peak optical depths are less than unity for all our sightlines except one (0431+206). Hence, an optically thin approximation to calculate the \HI~emission column density should be mostly valid. In column (9), we present the column density weighted harmonic mean spin temperatures ($\langle T_s \rangle$), defined as, $\langle T_s \rangle = \int T_B \thinspace dv / \int(1-e^{-\tau})\thinspace dv $, along our target sightlines. Columns (10) and (11) show the $\Delta_{90}$ of the absorption and the emission profiles, respectively, where $\Delta_{90}$ represents the width in \kms~containing the 90\% of total integral optical depth or emission.

Due to much higher column density and lower spin temperature, the CNM along a line-of-sight heavily biases the calculated harmonic mean spin temperature. For example, a line-of-sight having 90\% warm gas with a spin temperature of $\sim$ 8000 K and 10\% cold gas with a spin temperature of $\sim$ 100 K would produce a $\langle T_s \rangle \sim 900$ K. This indicates that any sightline with a high $\langle T_s \rangle$ must be dominated by the WNM. As can be seen from table~\ref{sample}, $\sim$ half of our sightlines have LOS averaged $\langle T_s \rangle > 500$ K, indicating the presence of a significant WNM along them. In Fig.~\ref{fig:ts} bottom panel, we present the per channel $\langle T_s \rangle (v)$ of our target sightlines. We calculate the per channel $\langle T_s \rangle (v)$ using $\langle T_s \rangle \thinspace (v) = T_B (v)/(1 - e^{-\tau (v)})$ ($T_B (v)$ and $\tau (v)$ are per channel values, integrated over the width of the spectral channel), where $T_B$ is the observed brightness temperature in LAB emission spectra as shown in the top panel of the figure. It should be mentioned here that the absorption spectra were resampled (as it has higher spectral resolution than the LAB spectra) to the velocities where the $T_B$ is measured in the LAB spectra. The solid red circles with error bars in the bottom panels represent the $\langle T_s \rangle$, where both the emission and the absorption spectra were detected with more than 3-sigma significance. In contrast, the blue points with up-arrows represent a 3-sigma lower limit to the $\langle T_s \rangle$. We note here that the $\langle T_s \rangle$ per spectral channel here cannot be directly associated with a specific component; rather, it is the column density weighted harmonic mean spin temperature of all the components that are overlapping in velocity corresponding to that specific channel. Hence, a $\langle T_s \rangle$ value in the stable CNM/WNM range is likely to indicate contributions dominantly from the corresponding phase, but $\langle T_s \rangle$ value in the `unstable' range can simply be a result of a mix of CNM and WNM contributing. However, the distribution of $T_B (v)$, $\tau(v)$ and/or $\langle T_s \rangle (v)$ contain more information, complementary to parameters extracted from Gaussian decomposition, of the distribution of the gas temperature than merely the line of sight average values of $T_s$. A detailed analysis of this is beyond the scope of the current work and we refer the readers to \citet{bhattacharjee2023} who has explored the same using realistic simulations of multiphase ISM.

\begin{figure}
\begin{center}
\resizebox{.45\textwidth}{!}{\includegraphics{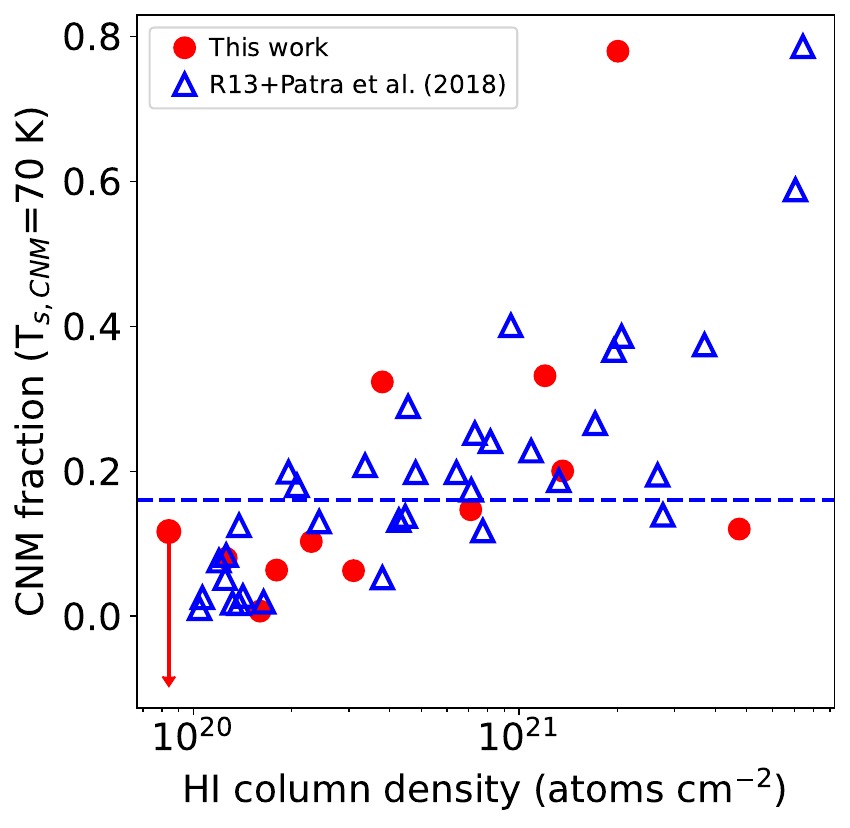}}
\end{center}
\caption{The lower limit on the CNM fraction along our observed sight lines. The red solid circles represent the current sample, whereas, the blue up triangles are taken from \citep{roy13a,patra18b}. The blue dashed line represents the median fraction of $\sim 50$\% for the full sample.}
\label{cnm_frac}
\end{figure}

\begin{figure}
\begin{center}
\resizebox{.45\textwidth}{!}{\includegraphics{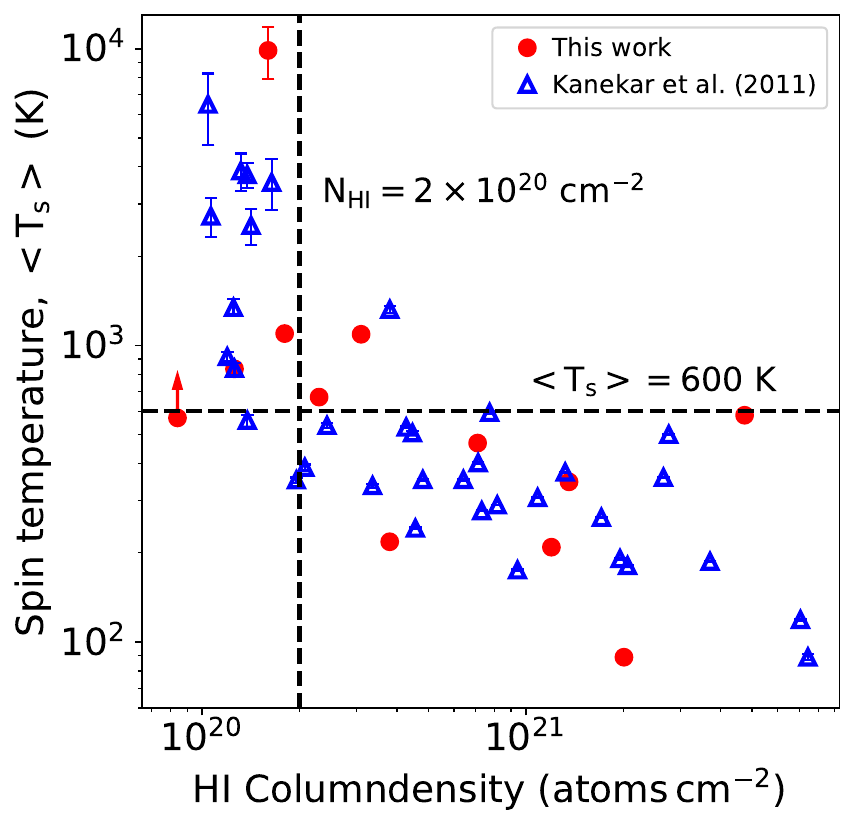}}
\end{center}
\caption{\HI~column density vs. $\langle T_s \rangle$ plot. The solid red circles are from this work whereas the blue up triangles are from our previous survey. As can be seen, there exist a threshold column density of $\sim 2 \times 10^{20}$ below which all the gas has high $\langle T_s \rangle$ indicating the dominance of WNM. See text for more details.}
\label{nh_ts}
\end{figure}

However, the high value of $\langle T_s \rangle$ in our sightlines only provides a qualitative account of CNM/WNM along our sightlines. A more quantitative lower limit of the CNM fraction can be estimated by using the absorption and emission line profiles. In the CNM, due to thermalization, the spin temperature approaches the kinetic temperature, which varies between 40-200 K. Given the optical depth spectrum, one can then calculate the CNM column density as

\begin{equation}
N_{HI,CNM} = 1.823 \times 10^{18} \times T_{s,CNM} \times \int \tau \thinspace dv
\label{eq1}
\end{equation}

\noindent where $T_{s,CNM}$ is the spin temperature of the CNM.

A lower limit to the observed \HI~column ($N_{HI,EM}$) density can be calculated using the emission spectra adopting an optically thin approximation. An optically thin approximation while calculating \HI~column density from the emission spectra guarantees a strict lower limit to the \HI~column density. At the same time, if one assumes all the gas is in CNM (with a spin temperature of $\sim$ 70 K), the calculated $N_{HI,CNM}$ would then represent a conservative upper limit to the observed \HI~column density. The ratio $N_{HI,CNM}/N_{HI,EM}$ then would provide a measure of the upper limit of the CNM fraction along a line-of-sight. In Fig.~\ref{cnm_frac}, we plot this ratio as a function of \HI~column density, $N_{HI,EM}$. For comparison, we also include the results for the 30 sources from our earlier studies \citep{roy13a,patra18b}. The median of the maximum CNM fraction for the current sample is found to be $\sim 12$\%, whereas the same for the entire sample is measured to be $\sim 16\%$ (blue dashed line in the figure).

It can also be seen from Fig.~\ref{cnm_frac} that the CNM fraction increases with increasing \HI~column density. It can be explained as with an increasing \HI~column density, self-shielding becomes more effective against background ionizing radiation, producing favorable conditions for CNM formation \citep{sofue2018,saha2018,murray15,murray2018,dempsey2022,basua2022}. This also suggests that there might exist a threshold \HI~column density below which it is tough to form CNM in the ISM. In fact, in an earlier study, \citet{kanekar11} has shown that there exists a threshold column density of $\sim 2 \times 10^{20} \ cm^{-2}$ below which no considerable amount of CNM has been found in galactic absorption spectra. In Fig.~\ref{nh_ts}, we plot the $\langle T_s \rangle$ as a function of \HI~column density for our sample (solid red circles). For completeness, we also plot the points from \citet{kanekar11} (blue up triangles). As can be seen from the figure, all the sightlines having column densities less than the threshold value (vertical dashed line) have spin temperatures $\geq 600$ K with a median of $\sim 2000$ K. On the other hand, the sightlines above the threshold column density have much lower $\langle T_s \rangle$ values, with a median of $\sim 350$ K. As discussed earlier, a $\langle T_s \rangle \sim 2000$ K means that, effectively, all the gas is in the WNM. Conversely, a $\langle T_s \rangle$ of $\sim$ 350 K signifies a considerable amount of cold gas along these sightlines.

\subsection{Muli-component Gaussian decomposition of absorption spectra}

However, the harmonic mean spin temperature is only an indicator of the possible dominance of CNM or WNM. It does not provide any quantitative account of the amount of CNM or WNM along a line-of-sight. To obtain the same, one needs more detailed modeling of the absorption/emission spectra. In the simplest description of the ISM, along any line-of-sight through the Galaxy, it can be thought to consist of multiple \HI~clouds with different physical properties. As a result, the observed absorption/emission spectrum will bear the signature of the physical properties of all the clouds. Theoretically, the CNM and the WNM in a thermalized cloud would coexist in pressure equilibrium \citep{field69b,wolfire95b,bialy2019}. Consequently, the absorption/emission spectra will have two Gaussian components with distinct widths. A kinetic temperature of $T_K$ would produce a width of (FWHM) $\Delta V = (T_K/21.855)^{\frac{1}{2}}$, $T_K$ is in Kelvin and $\Delta V$ in \kms. Under this circumstance, any line-of-sight hosting multiple \HI~cloud would produce an absorption profile consisting of several Gaussian components representing the physical properties of individual clouds.

The observed absorption/emission spectrum can then be decomposed into multiple Gaussian components to obtain the physical parameters of the individual clouds along the line-of-sight. This technique is widely used to understand the properties of the ISM in galaxies, both using emission and absorption spectra \citep{young96,young03,warren12,patra16,kanekar03a,roy13a,murray15,murray2018,basua2022,bhattacharjee2023}. However, the interpretation of the obtained parameters through modeling the observed \HI~spectra with multiple Gaussian components has several shortcomings. For example, in the ISM, the kinetic temperature does not decide the line width alone \citep[see, e.g.,][]{sridhar1994,goldreich95,kalberla2019,choudhuri2019,koley23}. The presence of non-thermal motions in the ISM primarily determines the final spectral widths. This, in turn, will only provide an upper limit to the kinetic temperature. Not only that, but the Gaussian components also do not form an orthogonal basis, and hence, any decomposition of the spectra can not be unique, especially in the presence of noise. This is a severe limitation as a complex spectrum can be fitted with different sets of Gaussian components with similar confidence. To avoid these shortcomings, there have been several efforts to fit the observed absorption spectra with more complicated model-based parametrization \citep{heiles03a,murray14}. However, these models, too, have several degeneracies and do not pose a higher degree of physical motivation than a simple Gaussian fit. Due to these reasons, a multi-component Gaussian fitting to the absorption spectra remains to date, one of the essential methods to investigate the neutral ISM in galaxies.

\begin{figure}
\begin{center}
\resizebox{.45\textwidth}{!}{\includegraphics{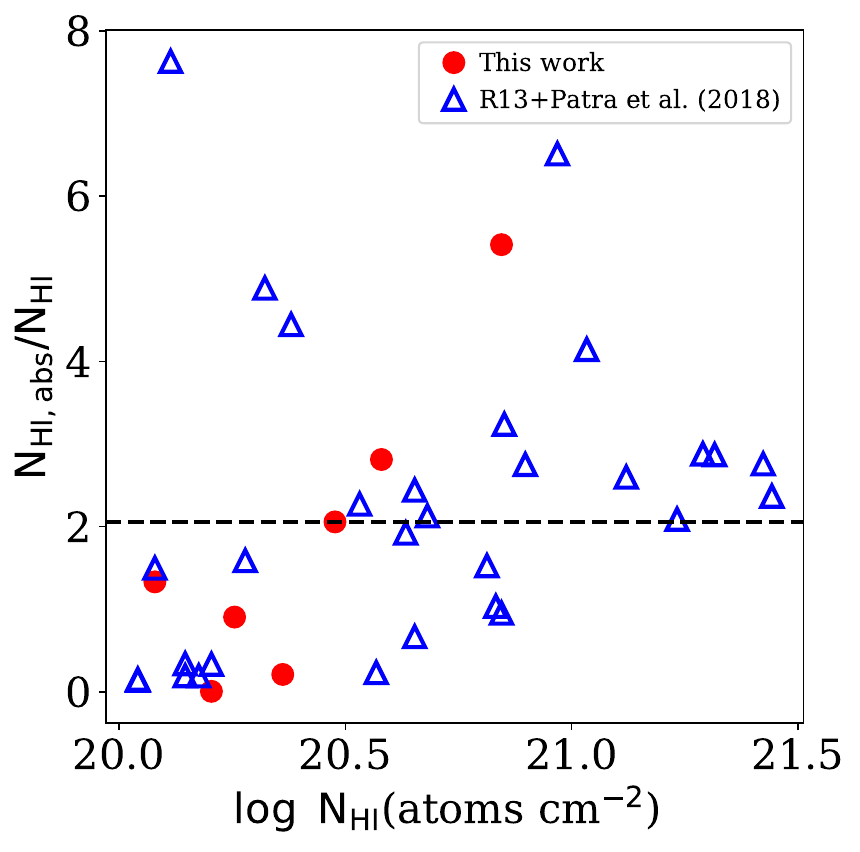}}
\end{center}
\caption{The ratio of $N_{HI,abs}$ to $N_{HI}$. The blue triangles are taken from \citep{roy13b}, whereas the red solid circles represent lines-of-sights presented in this work. The median value for $N_{HI,abs}/N_{HI}$ is found to be $\sim 2$. See text for more details.}
\label{nabs_nh}
\end{figure}

Hence, we fit the \HI~absorption spectra of our target sightlines using multiple Gaussian components. We decide the optimum number of Gaussian components required to fit a spectrum based on the reduced $\chi^2$ and the residuals. For a model well describing a spectrum, it is expected to result in a reduced $\chi^2$ close to unity and a noise-like residual. However, as the noise on our spectra is not uniform (see \S2), the residual (data - model) is not expected to follow a normal distribution unless normalized by the noise spectrum. Hence, we check for Gaussianity in the residual after normalizing it with the noise spectrum. We select a minimum number of Gaussian components to fit a spectrum, which results in a reduced $\chi^2$ close to unity and a normalized residual following a normal distribution. For four sightlines in our sample, i.e., 0431+206, 1924+334, 2007+404, and 2344+824, the optical depth spectra are found to be highly complex. As discussed earlier, there could be multiple sets of Gaussian components for these sightlines, which can produce equally reasonable models. Hence, exclude these sightlines and do not decompose them into multiple Gaussian components.

\begin{figure*}
\begin{center}
\begin{tabular}{ccc}
\resizebox{.32\textwidth}{!}{\includegraphics{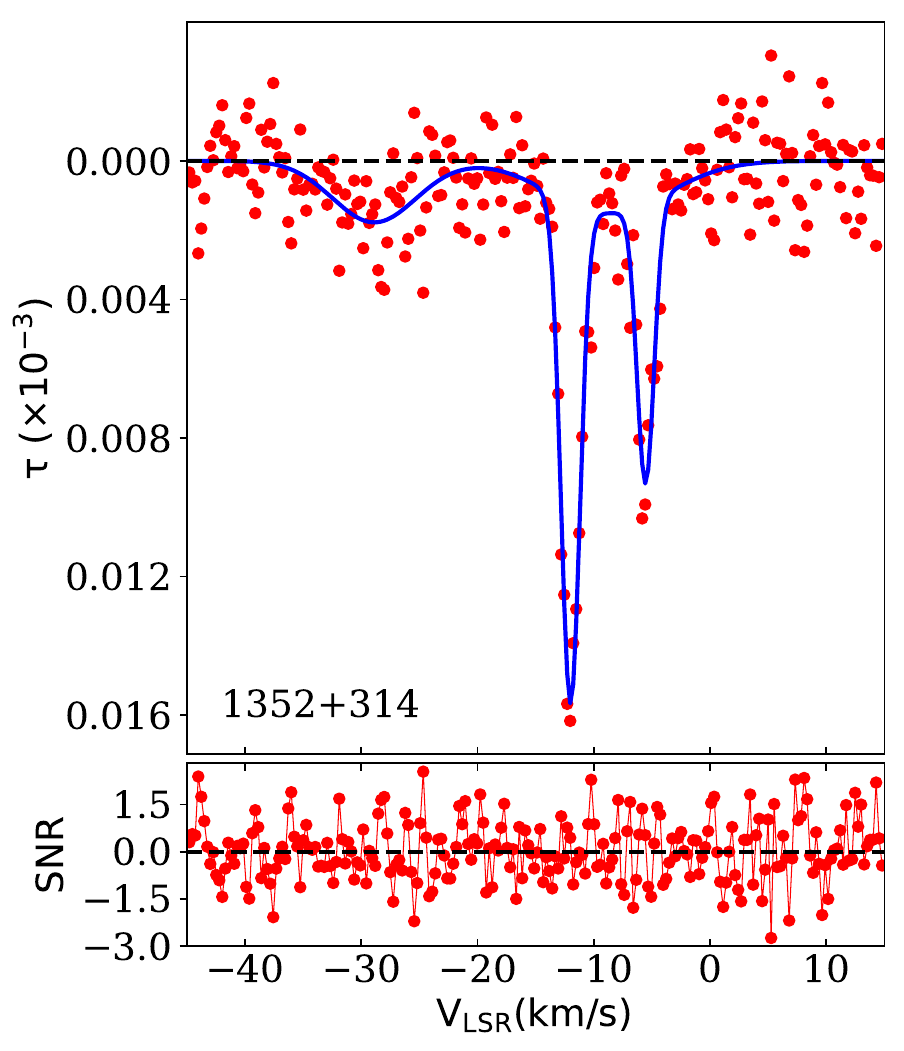}} &
\resizebox{.34\textwidth}{!}{\includegraphics{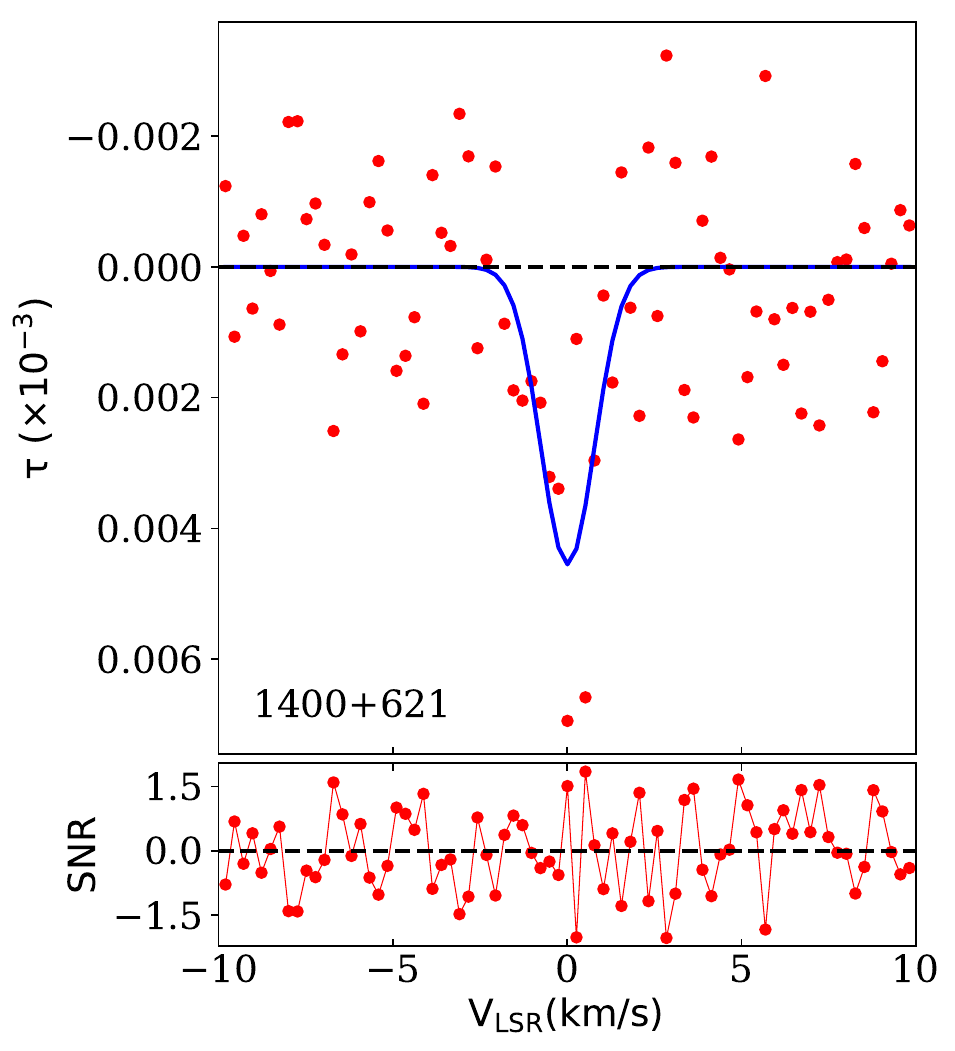}} &
\resizebox{.32\textwidth}{!}{\includegraphics{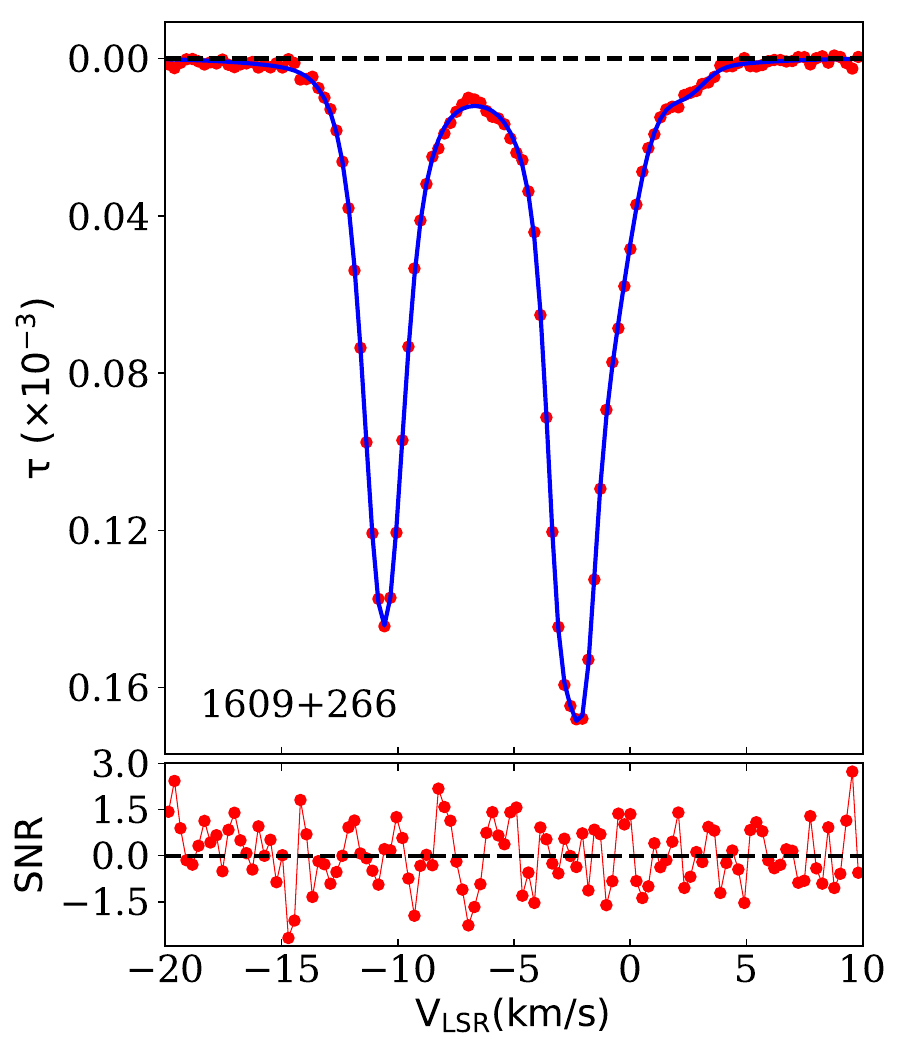}} \\
\resizebox{.33\textwidth}{!}{\includegraphics{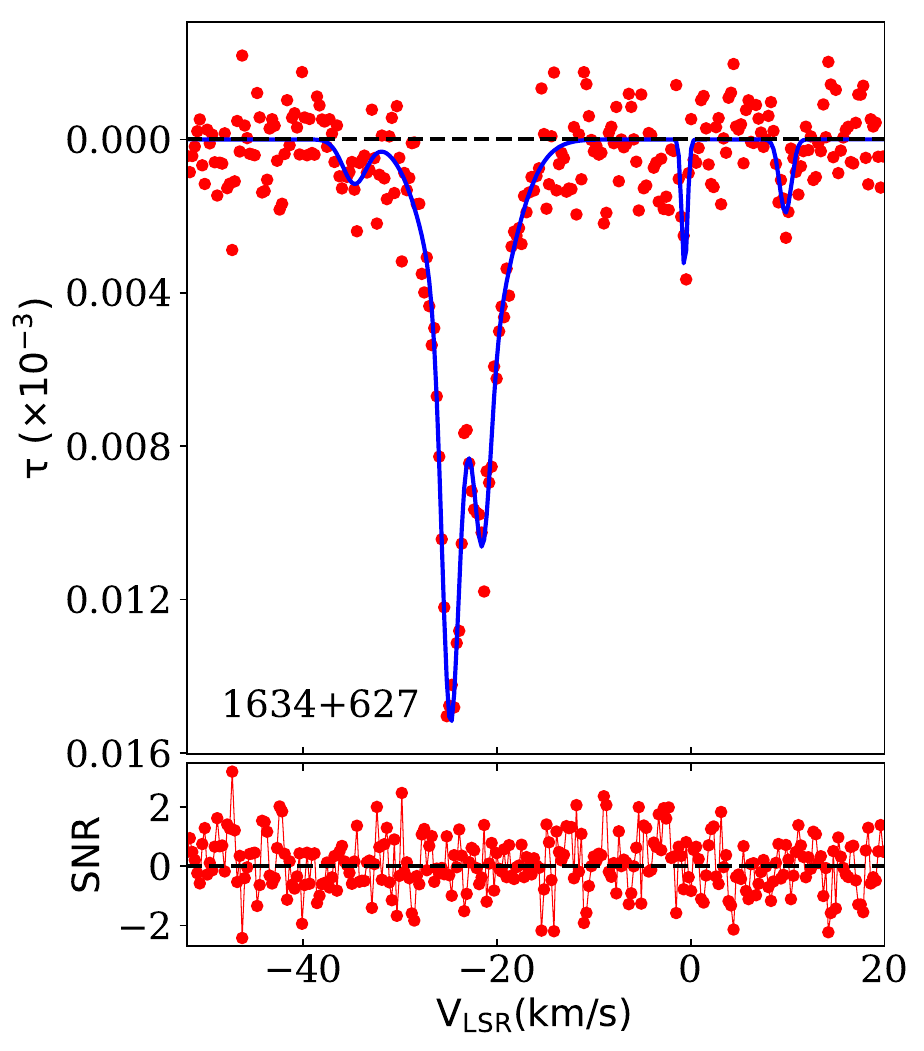}} &
\resizebox{.32\textwidth}{!}{\includegraphics{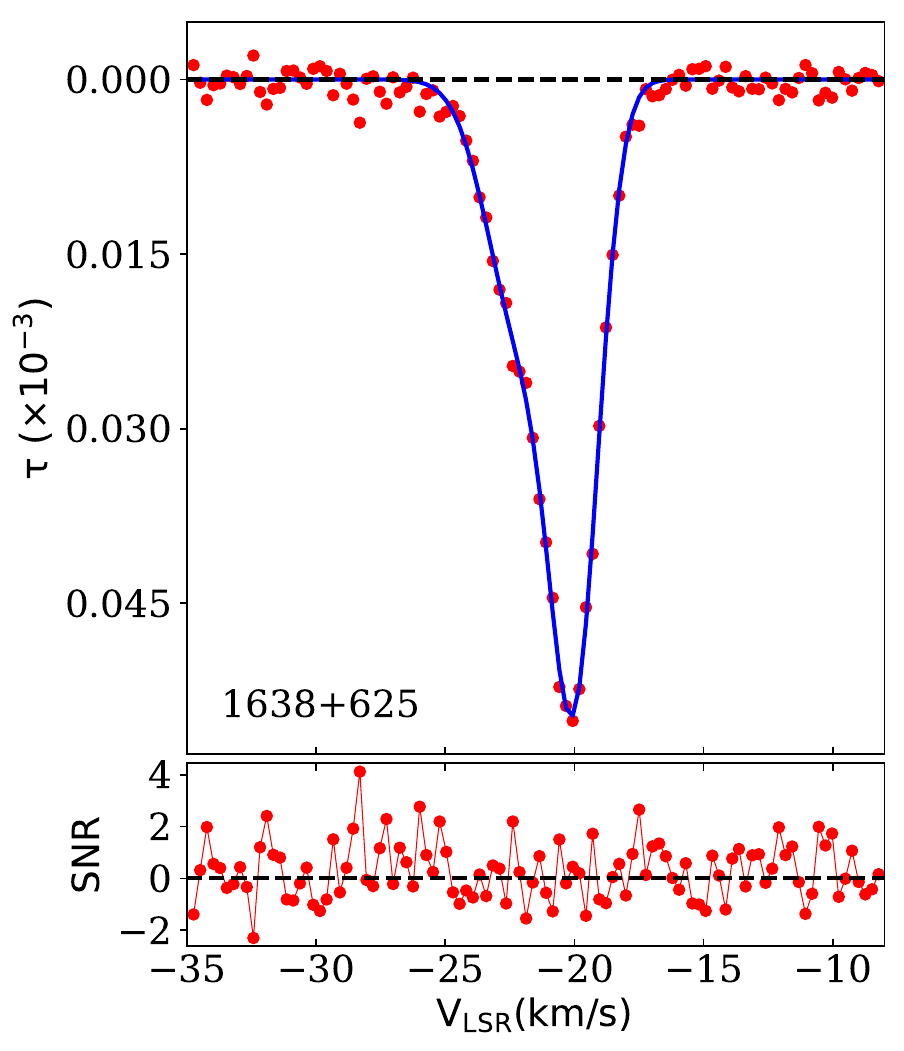}} &
\resizebox{.32\textwidth}{!}{\includegraphics{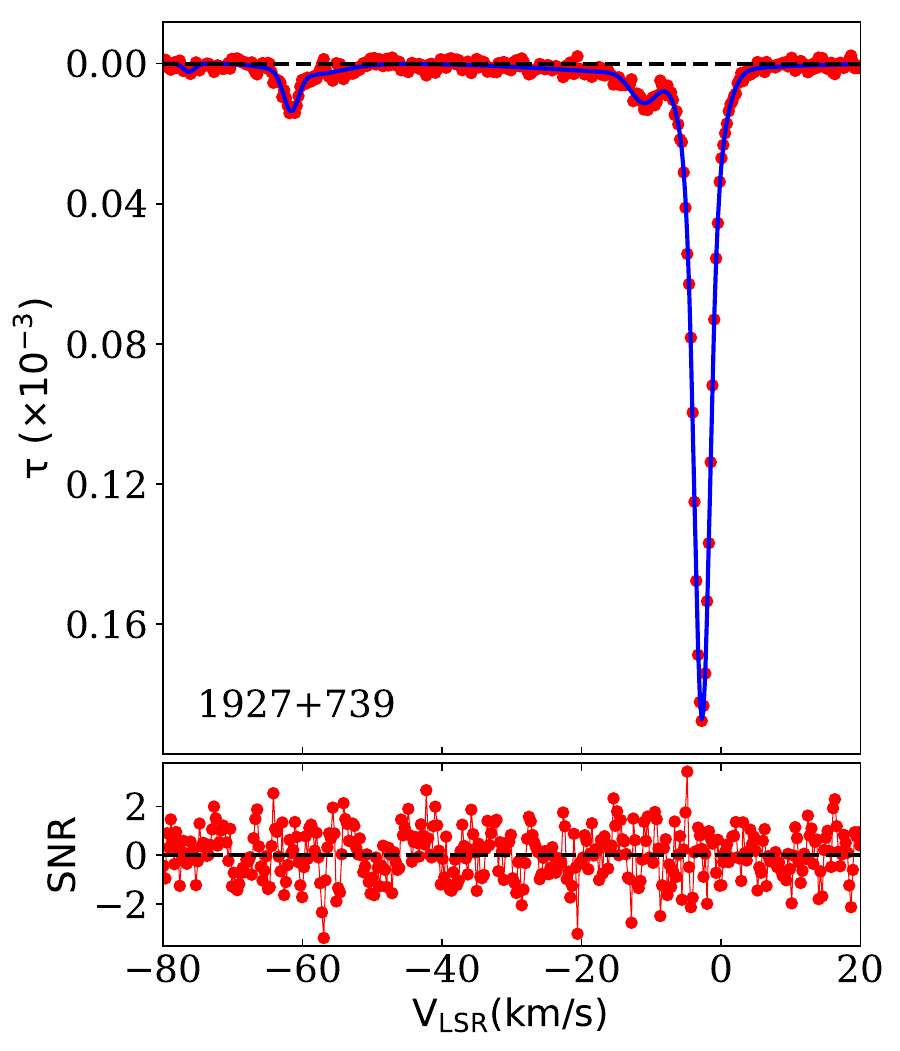}} \\
\resizebox{.34\textwidth}{!}{\includegraphics{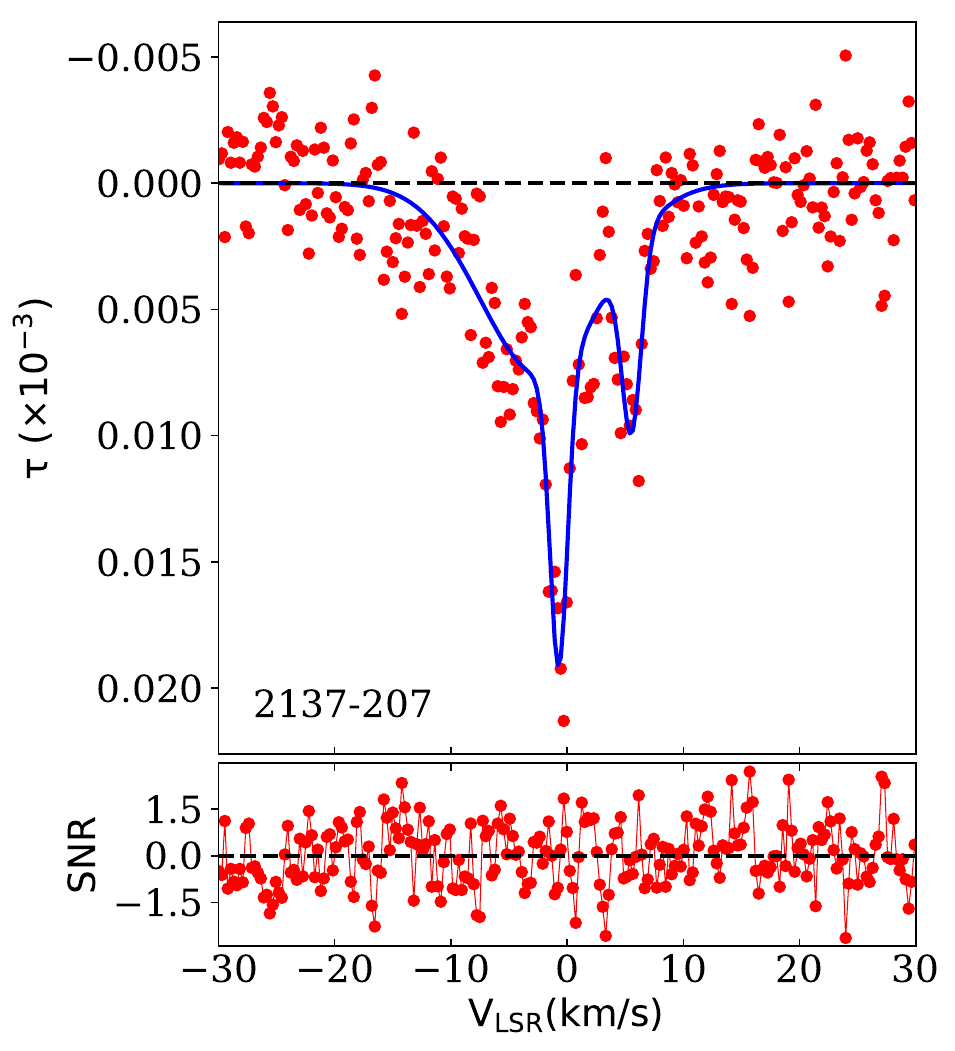}} \\
\end{tabular}
\end{center}
\caption{Gaussian decomposition of our absorption spectra. In the top panels, the solid red points represent the absorption spectra whereas the blue solid lines represent the fits. In the bottom panels, we show the residual of the fit normalised by the noise (See \S 3 for more details). As can be seen from the bottom panel, the normalised residuals look noise-like without any strong feature. See text for more details.}
\label{fig:gaussfit}
\end{figure*}

In Fig.~\ref{fig:gaussfit}, we show the Gaussian fittings to our absorption spectra. As can be seen from the figure, for our fits, the normalized residuals (bottom panel) resemble noise-like. However, to check Gaussianity, we performed a Kolmogorov-Smirnov rank-1 and Anderson-Darling tests with better than 99\% confidence. In table~\ref{tab:gaussfit}, we present the parameters of the fitted Gaussian components. The first column shows the name of the background source, and column (2) quotes the number of Gaussian components used to fit the spectra. Columns (3) and (4) show the peak optical depth and the location of the peak, respectively. Column (5) shows the $b$-parameter of the Gaussian component, and column (6) presents the doppler temperature. The last column in the table shows the achieved reduced $\chi^2$ values for individual spectra. As can be seen, the reduced $\chi^2$ for our spectra are close to unity, indicating good modeling of the spectra. The doppler temperature is calculated using $T_D = 21.855 \times \Delta v ^2$, where $\Delta v$ is the FWHM of the Gaussian component. As mentioned earlier, these Gaussian components can be regarded as representative of individual ISM phases, and hence, their properties can be further explored to investigate the ISM conditions in the Galaxy.

\begin{table*}
\caption{Properties of the decomposed Gaussian components.}
\centering
\begin{tabular}{c|c|c|c|c|c|c}
\hline
Name & $N_{comp}$ & $\tau_{peak}$ & $V_c$ & $b$ & $T_D$ & $\chi^2_{Red}$\\
     &      &               & (\kms)  & (\kms) & (K) & \\
\hline
1352+314 & 1 & 0.01446 $\pm$ 0.00064 & -12.01 $\pm$   0.04 &   1.16 $\pm$   0.06 & 80 $\pm$ 8 &  1.026 \\
         & 2 & 0.00807 $\pm$ 0.00067 &  -5.55 $\pm$   0.07 &   1.04 $\pm$   0.11 & 65 $\pm$ 13 &       \\
         & 3 & 0.00178 $\pm$ 0.00028 & -28.88 $\pm$   0.66 &   5.10 $\pm$   0.94 & 1573 $\pm$ 583 &       \\
         & 4 & 0.00151 $\pm$ 0.00036 &  -8.68 $\pm$   1.24 &   7.16 $\pm$   1.38 & 3106 $\pm$ 1201 &       \\
1400+621 & 1 & 0.00455 $\pm$ 0.00085 &   0.02 $\pm$   0.17 &   1.09 $\pm$   0.23 & 71 $\pm$ 30 &  0.863 \\
1609+266 & 1 & 0.09226 $\pm$ 0.00253 &  -1.87 $\pm$   0.03 &   2.08 $\pm$   0.03 & 262 $\pm$ 8 &  0.973 \\
         & 2 & 0.05487 $\pm$ 0.00548 &  -2.01 $\pm$   0.05 &   0.65 $\pm$   0.04 & 25 $\pm$ 3 &       \\
         & 3 & 0.00553 $\pm$ 0.00067 &   2.44 $\pm$   0.10 &   1.08 $\pm$   0.17 & 71 $\pm$ 21 &       \\
         & 4 & 0.01123 $\pm$ 0.00057 &  -5.67 $\pm$   0.37 &   7.21 $\pm$   0.24 & 3149 $\pm$ 212 &       \\
         & 5 & 0.04731 $\pm$ 0.00802 & -10.52 $\pm$   0.03 &   1.85 $\pm$   0.11 & 206 $\pm$ 24 &       \\
         & 6 & 0.06728 $\pm$ 0.00380 &  -3.01 $\pm$   0.05 &   0.73 $\pm$   0.04 & 32 $\pm$ 3 &       \\
         & 7 & 0.08992 $\pm$ 0.00813 & -10.61 $\pm$   0.01 &   1.02 $\pm$   0.03 & 62 $\pm$ 4 &       \\
1634+627 & 1 & 0.00948 $\pm$ 0.00117 & -24.84 $\pm$   0.06 &   1.20 $\pm$   0.13 & 87 $\pm$ 19 &  0.963 \\
         & 2 & 0.00436 $\pm$ 0.00123 & -21.45 $\pm$   0.13 &   1.14 $\pm$   0.30 & 78 $\pm$ 41 &       \\
         & 3 & 0.00339 $\pm$ 0.00080 &  -0.63 $\pm$   0.08 &   0.42 $\pm$   0.11 & 10 $\pm$ 5 &       \\
         & 4 & 0.00192 $\pm$ 0.00055 &   9.78 $\pm$   0.20 &   0.87 $\pm$   0.29 & 45 $\pm$ 29 &       \\
         & 5 & 0.00674 $\pm$ 0.00147 & -22.84 $\pm$   0.21 &   4.93 $\pm$   0.47 & 1470 $\pm$ 278 &       \\
         & 6 & 0.00114 $\pm$ 0.00038 & -34.68 $\pm$   0.48 &   1.75 $\pm$   0.69 & 184 $\pm$ 144 &       \\
1638+625 & 1 & 0.04821 $\pm$ 0.00347 & -19.99 $\pm$   0.05 &   1.34 $\pm$   0.04 & 108 $\pm$ 6 &  1.088 \\
         & 2 & 0.02119 $\pm$ 0.00164 & -22.06 $\pm$   0.23 &   1.85 $\pm$   0.17 & 207 $\pm$ 38 &       \\
1927+739 & 1 & 0.14169 $\pm$ 0.00396 &  -2.73 $\pm$   0.01 &   1.51 $\pm$   0.02 & 137 $\pm$ 4 &  1.022 \\
         & 2 & 0.04373 $\pm$ 0.00406 &  -2.62 $\pm$   0.04 &   3.27 $\pm$   0.13 & 646 $\pm$ 51 &       \\
         & 3 & 0.01091 $\pm$ 0.00067 & -61.67 $\pm$   0.07 &   1.52 $\pm$   0.12 & 140 $\pm$ 21 &       \\
         & 4 & 0.00323 $\pm$ 0.00044 & -59.05 $\pm$   0.61 &   6.38 $\pm$   0.67 & 2464 $\pm$ 519 &       \\
         & 5 & 0.00253 $\pm$ 0.00038 & -10.94 $\pm$   1.33 &  18.63 $\pm$   1.82 & 21040 $\pm$ 4100 &       \\
         & 6 & 0.00873 $\pm$ 0.00056 & -11.00 $\pm$   0.13 &   2.74 $\pm$   0.22 & 455 $\pm$ 71 &       \\
         & 7 & 0.00237 $\pm$ 0.00062 & -76.28 $\pm$   0.24 &   1.13 $\pm$   0.34 & 76 $\pm$ 46 &       \\
2137-207 & 1 & 0.01183 $\pm$ 0.00127 &  -0.72 $\pm$   0.09 &   1.08 $\pm$   0.15 & 70 $\pm$ 19 &  1.069 \\
         & 2 & 0.00735 $\pm$ 0.00116 &   5.55 $\pm$   0.14 &   1.19 $\pm$   0.24 & 85 $\pm$ 34 &       \\
         & 3 & 0.00761 $\pm$ 0.00063 &  -2.20 $\pm$   0.47 &   7.44 $\pm$   0.62 & 3355 $\pm$ 557 &       \\
\hline
\end{tabular}
\label{tab:gaussfit}
\end{table*}

\begin{figure}
\begin{center}
\resizebox{.45\textwidth}{!}{\includegraphics{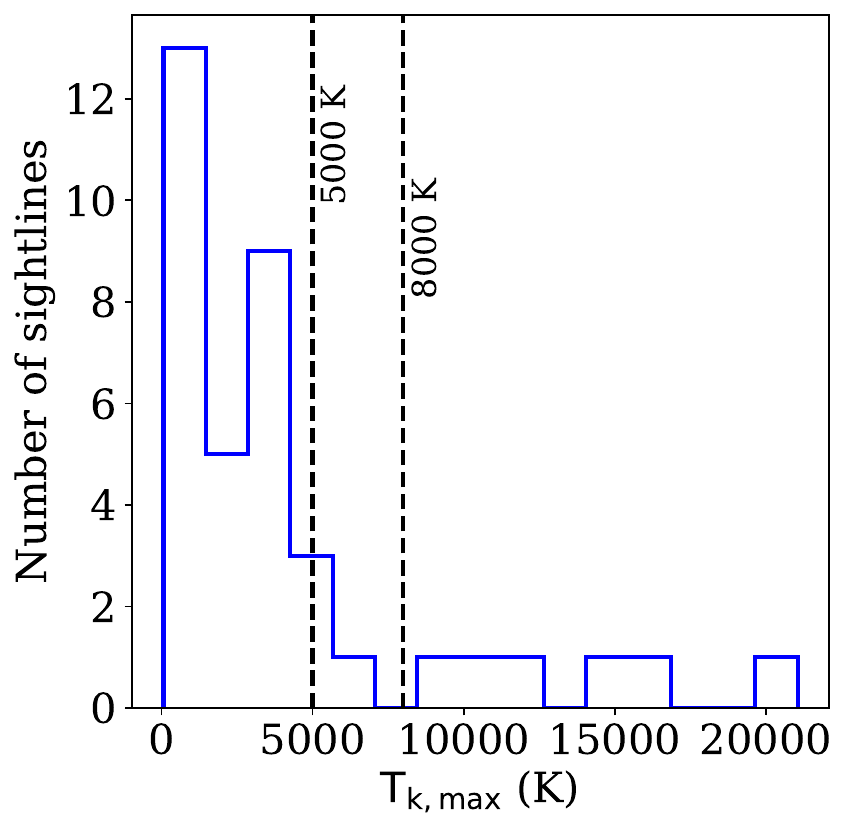}}
\end{center}
\caption{The histogram of maximum kinetic temperatures (of all the clouds) along different sightlines. The maximum kinetic temperatures are chosen amongst all the decomposed Gaussian components along a line of sight. This histogram includes sightlines from this study and \citet{roy13b}. We found 16 sightlines for which the maximum kinetic temperatures are considerably lower than the stable WNM. We use these sightlines to test a strict two-phase model. See the text for more details.}
\label{kinetic}
\end{figure}

In table~\ref{tab:gaussfit}, we list 30 Gaussian components from our absorption spectra. We assume that these Gaussian components represent different phases of the ISM in the Galaxy. Along with these components, we also include 214 Gaussian components from our previous studies \citep{roy13b,patra18c}. This makes the total number of Gaussian components 244, which we use to investigate the properties of the ISM. Using the FWHM of individual Gaussian components, we estimate the doppler temperature, $T_D$ (or $T_{k,max}$), and estimate the fraction of the clouds having $T_D$ consistent with that of CNM ($T_{k,max} \textless 500$K), WNM ($T_{k,max} \textgreater 5000$K) or the unstable medium, UNM ( $500 \textless T_{k,max} \textless 5000$). We find that for the full sample, $\sim$ 72\%, 24\%, and 4\% of the total components have $T_D$  consistent with that of the CNM, UNM, and WNM, respectively. We emphasize that these fractions are not the gas fractions along a line-of-sight but are based solely on the number of components. To estimate the fractional column densities in these ISM phases, a knowledge of the spin temperature (\ts) of the individual components is essential. However, \ts~is not a directly measurable quantity in an absorption spectrum, and estimation of which requires complicated modeling, which very often produces degenerate solutions for moderately complex spectra \citep{roy13b}.

However, the doppler temperature can be used as an upper limit to the spin temperature. An assumption of $T_s \approx T_{k,max}$ then can be used to calculate an upper limit to the column density in the absorption spectra, $N_{HI,abs}$. Where,

\begin{equation}
\begin{split}
N_{HI,abs} &= 1.823 \times 10^{18} \times T_{k,max} \times \int (1 - e^{-\tau(v)}) \thinspace dv\\
           &\simeq 1.823 \times 10^{18} \times T_{k,max} \times \int \tau(v) \thinspace dv
\end{split}
\label{eq2}
\end{equation}

\noindent Here we assume our sightlines to be optically thin with $\tau (v) \textless 0.5$ (which is true for all the sightlines with a possible exception along 0431+206).

For an optically thin cloud, column density can be given in terms of observed brightness temperature, $T_B$ as,

\begin{equation}
N_{HI} = 1.823 \times 10^{18} \int T_B (v) \thinspace dv
\end{equation}

On the other hand, the emission spectra are often used to calculate the column density along a line-of-sight, $N_{HI}$, employing an optically thin approximation. This, $N_{HI}$, represents a lower limit to the line-of-sight column density. Ideally, for a thermalized cloud, in the absence of any non-thermal broadening and low optical depth (optically thin approximation holds), $N_{HI,abs}$ should match $N_{HI}$ closely. However, due to the presence of non-thermal broadening in the ISM \citep{audit05,hennebelle2012,choudhuri2019,kalberla2019,koley19}, $T_{k,max}$ is expected to be larger than \ts~resulting $N_{HI,abs}$ to be higher than $N_{HI}$. In this sense, the ratio $N_{HI,abs}/N_{HI}$ can indicate the non-thermal broadening along a line-of-sight. For example, using Eq.~\ref{eq1}, and Eq.~\ref{eq2} one can obtain,  

\begin{equation}
N_{HI,abs} = N_{HI,CNM} \times \frac{T_{k,max}}{T_{s,CNM}}.
\label{eq3}
\end{equation}

In the above equation, $N_{HI, abs}$ is proportional to the ratio $T_{k,max}/T_{s,CNM}$, which is a measure of the non-thermal broadening. Thus, $N_{HI,abs}/N_{HI}$ along any line-of-sight can be used as an indicator of the non-thermal broadening. In Fig.~\ref{nabs_nh} we plot $N_{HI,abs}/N_{HI}$ as a function of $N_{HI}$. If there is no non-thermal broadening along our sightlines, an $N_{HI,abs}/N_{HI}$ value close to unity is expected. However, as can be seen from the figure, there are several sightlines for which $N_{HI,abs}/N_{HI} \textgreater 1$ with an overall median value of $\sim 2.1$. This indicates the existence of a significant non-thermal broadening along our sightlines.

\begin{figure}
\begin{center}
\resizebox{.45\textwidth}{!}{\includegraphics{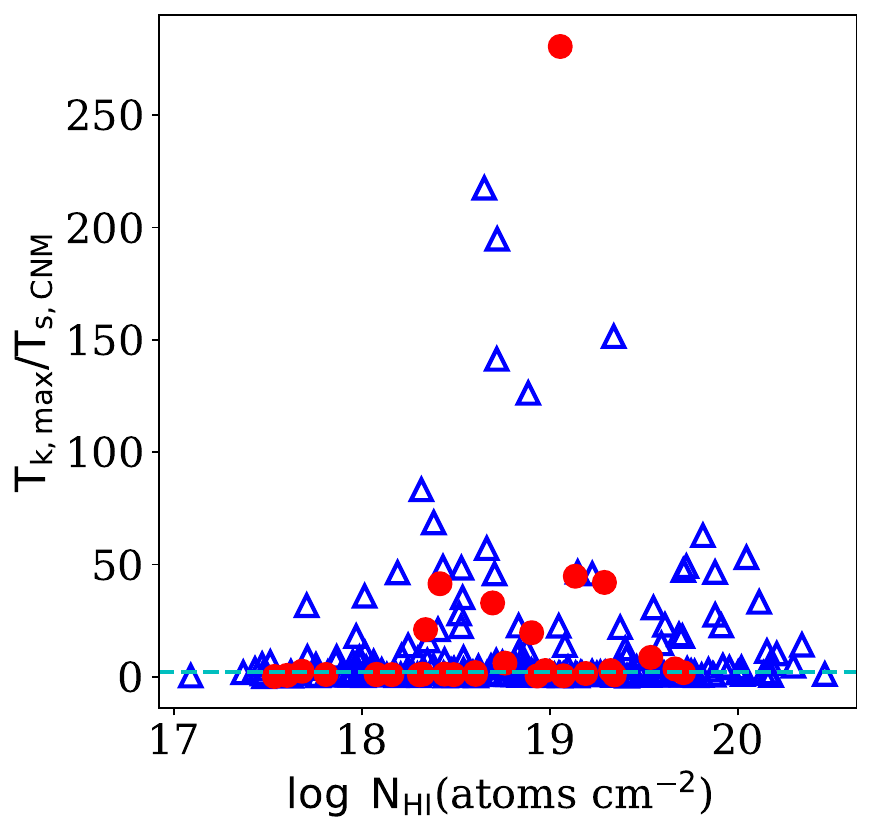}}
\end{center}
\caption{The ratio of the maximum kinetic temperature, $T_{k,max}$ to the spin temperature of the CNM, $T_{s,CNM}$ of the decomposed Gaussian components. The red points are from this study, whereas the blue triangles are from our earlier papers \citep{roy13b,patra18c}. This ratio acts as an indicator of the non-thermal broadening of the components. The median ratio is found to be 2.1, which is the same as the median of $N_{HI,abs}/N_{HI}$ (Fig.~\ref{nabs_nh}). See the text for more details.}
\label{td_ts}
\end{figure}

Further, in Fig.~\ref{td_ts}, we plot $T_{k,max}/T_{s,CNM}$ as a function of the column density of individual Gaussian components (assuming a spin temperature of 70 K). Ideally, as this ratio implies the amount of non-thermal broadening, we expect a similar result as found in Fig.~\ref{nabs_nh}. Interestingly, we found here also many components with $T_{k,max}/T_{s,CNM} > 1$, with a median value of $\sim 2.1$. This indicates the consistency of the parameter $N_{HI,abs}/N_{HI}$ for measuring non-thermal broadening and our assumption of $T_{s,CNM} \sim 70$ K. 

Using the parameters of the decomposed Gaussian components, we investigate if a purely two-component model of the ISM consisting of only CNM and WNM is consistent with our observations. We first identify the sightlines where the maximum kinetic temperature, $T_{k,max}$, is much lower than classical WNM temperatures. In Fig.~\ref{kinetic}, we plot the histogram of the maximum kinetic temperatures along our sightlines. We find 16 sightlines from the previous sample, and four sightlines in the current sample have kinetic temperatures much lower than that of the WNM. If a two-phase model is prevalent in these sightlines, all the gas along them can be considered to have originated from the CNM. Then, a higher $T_{k,max}$ can be attributed to a non-thermal broadening. In that case, for this gas, a maximum CNM spin temperature of $T_{s,CNM} = 200$ K can be adopted to calculate a maximum column density ($N_{H,CNM,max}$) expected to be detected in absorption.  

If there is no WNM component along these sightlines, this $N_{H,CNM,max}$ should always be higher than the column density estimated from the emission spectra, $N_H$. However, we find that for all the selected sightlines, $N_H > N_{H,CNM,max}$. Nonetheless, WNM may exist along these sightlines, but they were not detected in our absorption spectra due to a lack of sensitivity. To check the same, we calculate at what significance we should have detected this WNM gas given our observing parameters. We conservatively use a spin temperature of 5000 K for WNM and a line width of $\Delta V_{90}^{em}$ to calculate the expected peak optical depth. We use this peak optical depth and the RMS optical depth of our observations to calculate the significance at which the WNM should have been detected. We find that for 13 sightlines out of 16 from the previous sample and all four from the current sample, WNM should have been detected with more than 3-sigma significance. This emphasizes the non-existence of a strict two-phase medium along these sightlines, given our estimates are reasonably conservative.

To investigate the distribution of the spin temperatures in the Galaxy and see if a two-temperature model can explain our observation, we build a model of the Milky Way and simulate the ISM conditions along our observing sightlines. The details of these simulations are presented in the Appendix. We find that a simple two-temperature model could not explain the observed integral optical depths along our observed sightlines, and it suggests a more complex spin temperature distribution in the Galaxy. 

\section{Conclusion}

Using the Westerbork Synthesis Radio Telescope in frequency switching mode, we observed 12 sightlines against background quasars to detect Galactic \HI~in absorption. We detect \HI~in absorption along all our sightlines except one, along 0431+206. Utilizing 12 hours observing time per source, we achieve an excellent optical depth RMS of $\sim 1 - 2 \times 10^{-3}$ per 0.26 \kms~channel width. An optical depth sensitivity that we accomplish here is essential for detecting low optical depth WNM in the absorption spectra.

We further use the absorption spectra obtained along our target sightlines to asses the properties of the neutral ISM in the Galaxy. Using the spectra, we calculate the intensity weighted harmonic mean spin temperature, $\langle T_s \rangle$, and find that $\sim 50\%$ of our sightlines have $\langle T_s \rangle \textgreater 500$ K indicating the existence of a high WNM fraction along these sightlines.

Using the absorption spectra, we also calculate a conservative upper limit to the CNM fraction in our sightlines. Counting all the gas in CNM, a maximum CNM column density is calculated by assuming a maximum CNM spin temperature of 70 K. Comparing this CNM column density with the observed column density from emission spectra, we estimate a strict upper limit on the CNM fraction along our sightlines. We find a median CNM fraction of $\sim 0.12$ for our current sample, and this fraction increases to $\sim 0.16$ for the total sample, including 30 sightlines from \citet{roy13b} and \citet{patra18b}.

The CNM fraction along any sightline is expected to depend on the \HI~column density as it helps in self-shielding. Using our previous observations, \citet{kanekar11} have shown that there exists a threshold column density of $2 \times 10^{20}$ \acc~below which, the ISM is dominated by WNM as indicated by a high $\langle T_s \rangle$. In the current sample, with the addition of 12 sightlines, we observe a similar trend and find that the threshold column density, as found by \citet{kanekar11}, continues to hold good for the Galaxy.

Assuming the most straightforward description of the ISM, we decompose the absorption spectra into multiple Gaussian components to identify different phases of the ISM. We use the parameters of these Gaussian components to examine the properties of the ISM phases. We find that the maximum kinetic temperature of $\sim 72\%$ of the total components (in our full sample) consistent with CNM temperature while only $\sim 4$ of the components show the kinetic temperature of WNM gas. The rest of the detected components have a maximum kinetic temperature in the unstable medium or UNM. Further, assuming the kinetic temperature of each Gaussian component to be the upper limit of the spin temperature, we calculate the line-of-sight column density, $N_{HI,abs}$. In the absence of any non-thermal broadening in a thermalized cloud, this $N_{HI,abs}$ expected to match the $N_{HI}$, the column density measured using \HI~emission spectrum. Thus, the ratio $N_{HI,abs}/N_{HI}$ serves as an indicator of the presence of non-thermal broadening in the ISM. We find that, for our full sample, there is significant non-thermal broadening with a median $N_{HI,abs}/N_{HI} \sim 2$.

We further use the widths of the Gaussian components to calculate the maximum kinetic temperature along a line-of-sight and identify the lines-of-sight where the maximum kinetic temperatures are much lower than the classical WNM temperatures. Assuming a classical two-phase model, we calculate the maximum possible CNM column density along these sightlines by adopting a maximum spin temperature of 200 K for CNM. Using the observed \HI~column density in emission, we then calculate the minimum WNM column density along each line-of-sight and the SNR at which this WNM gas is supposed to be detected in our observations. We find that, if all the gas were to be in a strict two-phase ISM, at least 13 out of 16 sightlines in our earlier sample \citep{roy13b} and all four sighlines in our current sample would have detected WNM in them. 

We explored a two-temperature model to describe the \HI~disk in the Galaxy and investigate if such a description can explain the observed total optical depth distribution along our observed sight lines. We find that the two-temperature model fails to reproduce our observed integral optical depths and suggests a more complex spin temperature distribution in the Galaxy.

\section{Data availability}
The GMRT/WSRT raw data used in this paper are publicly available in the GMRT/WSRT archive. However, the data products can be made available upon a reasonable request.

\section*{Acknowledgements}
We are grateful to Harvey Liszt for providing N.R. with his simulation results. We are grateful to Nissim Kanekar and Jayaram N. Chengalur for their valuable discussions and suggestions, which helps to improve this paper immensely. NNP acknowledges support from the Science and Engineering Research Board (SERB) of the Department of Science and Technology (DST), Government of India, through the Startup Research Grant (SRG) no - SRG/2022/000917.

\section*{Appendix}



\subsection*{Spin temperature distribution in the disk of the Galaxy}

To investigate the spin temperature distribution in the Galaxy and check the validity of a simple two-temperature disk, we simulate different quantities along our sightlines and compare them to the observation. We assume a physically motivated two-temperature model for the \HI~disk in the Galaxy, a cold, thin disk with a low spin temperature and a warm disk with a higher spin temperature. We first build a three-dimensional density model of the \HI~disk to simulate different measurable quantities. We adopt the gas disk model from \citet{mcmillan2017}, which fits observational constraints and is consistent with theoretical expectations. The \HI~disk of the Milky Way can be represented by a double exponential,

\begin{equation}
    \rho_d(R,z) = \frac{\Sigma_0}{4z_d} \exp{\left( -\frac{R_m}{R} - \frac{R}{R_d}\right)} \ sech^2 (z/2z_d)
    \label{eq:rhoden}
\end{equation}

\begin{figure}
\begin{center}
\resizebox{.45\textwidth}{!}{\includegraphics{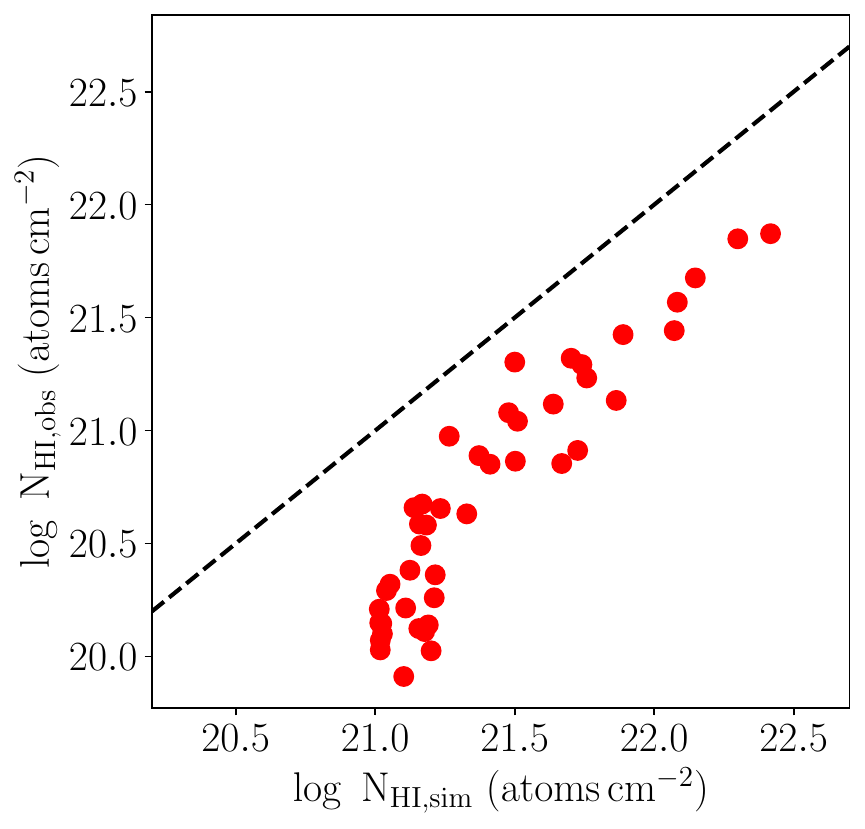}}
\end{center}
\caption{Comparison of simulated and observed column densities along our lines of sight. The horizontal and vertical axes represent simulated and observed column densities, respectively. The red circles represent column densities without optimization, i.e., considering the \HI~disk model parameters as quoted by \citet{mcmillan2017} ($\Sigma_0=53.1 \ M_{\odot} \thinspace pc^{-2}$, $R_d = 7$ kpc, $R_m=4$ kpc). The cyan squares indicate model parameters optimized for individual sightlines. The black dashed line shows the 1:1 line. As can be seen, the observed and simulated column density matches well for individual optimization, whereas, for the non-optimized case, they fail to produce the observed column densities along all the sight lines. See the text for more details.}
\label{colden_comp}
\end{figure}

\noindent This represents an exponential disk with a central density of $\Sigma_0$ and a scale length of $R_d$. The disk has a central hole with an associated scale length of $R_m$. $z_d$ represents a constant vertical scale height with radius. For the \HI~disk, \citet{mcmillan2017} found the values for $\Sigma_0$, $R_d$, $R_m$, and $z_d$ to be, $53.1 \ M_{\odot} \thinspace pc^{-2}$, $7$ kpc, 4 kpc, and 85 pc, respectively. The density distribution in the vertical direction was considered to follow a $sech^2$ profile with a constant scale height, $z_d$, throughout. However, the scale height in galaxies is found to increase with radius \citep{patra2020,patra2019}, and hence, for the Galaxy, we adopt an exponentially increasing scale height, given by \citep{kalberla2009},

\begin{equation}
    z (R) = z_0 \exp{R/R_z}
    \label{eq:sclh}
\end{equation}

\noindent where, $z_0 = 150$ pc, and $R_z = 9.8$ kpc \citep[see,][for more details]{kalberla2009}. 

We calculate the \HI~column densities along our sightlines using this description of the atomic disk and compare them with observation in Fig.~\ref{colden_comp}. As can be seen from the figure, an azimuthally symmetric model of \citet{mcmillan2017} systematically produces lower column densities (solid red circles) than what is observed. We note that we have only 40 sight lines for which we compare the column densities. \citet{mcmillan2017} optimized the model parameters using global data of much larger size.

Nonetheless, we explore the possibility of different model parameters for our observations than what is used by \citet{mcmillan2017}. We also consider that these parameters can be direction-dependent, i.e., in different directions, the disk can be described by different parameters (non-symmetric model). With these considerations, we perform an MCMC to find optimized galaxy parameters in each sightline matching the observed column density. For each sight line, we construct 10,000 uniformly sampled parameter sets and calculate the column densities for each of them. Consequently, we find the best parameter set by a $\chi^2$ minimization. For MCMC, we adopt a range for $\Sigma_0: 5-50 \ M_{\odot} \thinspace pc^{-2}$, $R_d: 5-9$ kpc, and $R_m: 2-6$ kpc. We use a flaring scale height as described by Eq.~\ref{eq:sclh} \citep{kalberla2009}, but do not vary its parameters for MCMC.


With this optimization, we get the best-fit galaxy density model for each sight line. Using these models, we then test if a two-temperature disk can reproduce observed integral optical depth along our sightlines. We assume that the \HI~disk consists of two disks, i.e., a cold, thin disk with a low spin temperature and a warm, thick disk with a high spin temperature. We also assume that there exists a characteristic height, $h_{spin}$, which separates these two disks. It is well known that the ISM in the Galaxy is predominantly found in one of the two stable phases, i.e., CNM and WNM \citep[see, e.g., ][]{wolfire95b}. As the CNM is dynamically cold, they are expected to settle to a thin disk close to the midplane. On the other hand, the WNM, having a much higher kinetic temperature, is expected to extend much beyond the midplane. Our two-temperature model is motivated by these expectations, which we probe using our observations.

\begin{figure}
\begin{center}
\resizebox{.45\textwidth}{!}{\includegraphics{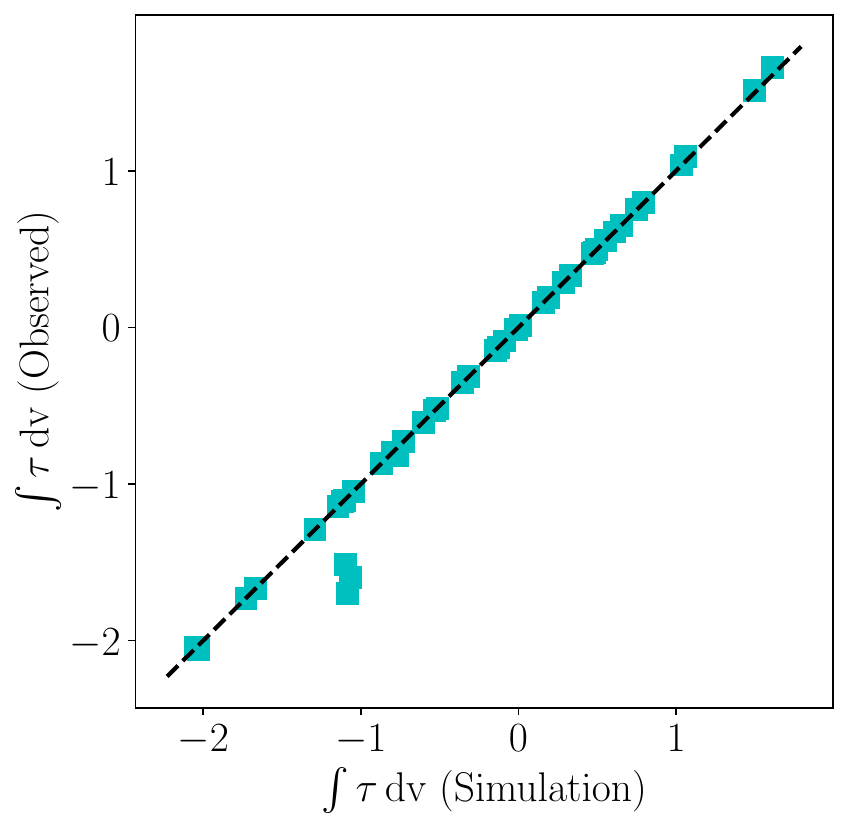}}
\end{center}
\caption{Optimized $\tau_{int}$ for a two-temperature model. The cyan squares represent $\tau_{int}$ for optimized models along individual sightlines. The black dashed line indicates the 1:1 line. As can be seen, the optimized integral optical depth values match the observation reasonably well. See the text for more details.}
\label{fig:tau_comp}
\end{figure}

We assume that below the height $h_{spin}$, the spin temperature assumes values between $50-500$ K, and above the height, it can vary between $1000-10000$ K. With the density and the spin temperature model, we calculate the integral optical depth along all our sightlines, which then can be compared to the observation. We adopt a similar MCMC approach to find the best spin temperatures and $h_{spin}$. For every sightline, we construct 10,000 parameter sets ($T_{s, CNM}: 50-500$ K, $T_{s, WNM}: 1000-10000$ K, $h_{spin}: 10-200$ pc) and calculate integral optical depths for all of them. We then use a $\chi^2$ minimization method to find the best values for $T_{s,CNM}$, $T_{s, WNM}$, and $h_{spin}$ which explain the observed integral optical depth. In Fig.~\ref{fig:tau_comp}, we show the optimized integral optical depth for our sight lines. As can be seen, with adjustable temperatures and $h_{spin}$, the optimized $\tau_{int}$ values explain the observations very well. However, there are three points (around $\tau_{int} \sim -1$) for which the optimized $\tau_{int} \sim -1$ do not match the observed value. We note that along these sightlines, the integrated number density of the $N_{HI}$ is so high that lower optical depths cannot be produced by altering the spin temperatures or the height $h_{spin}$ within our chosen limit (physically motivated from observation). We conclude that, along these sightlines, the volume filling factor must be much less than one, and a uniform density distribution cannot accurately model the ISM. A detailed modeling is required to capture the intricate details of the ISM structure. Further, our results indicate that a symmetric model with global spin temperatures is not favored, and a substantial variation in the model parameter is expected along different sight lines.

\begin{figure*}
\begin{center}
\resizebox{1\textwidth}{!}{\includegraphics{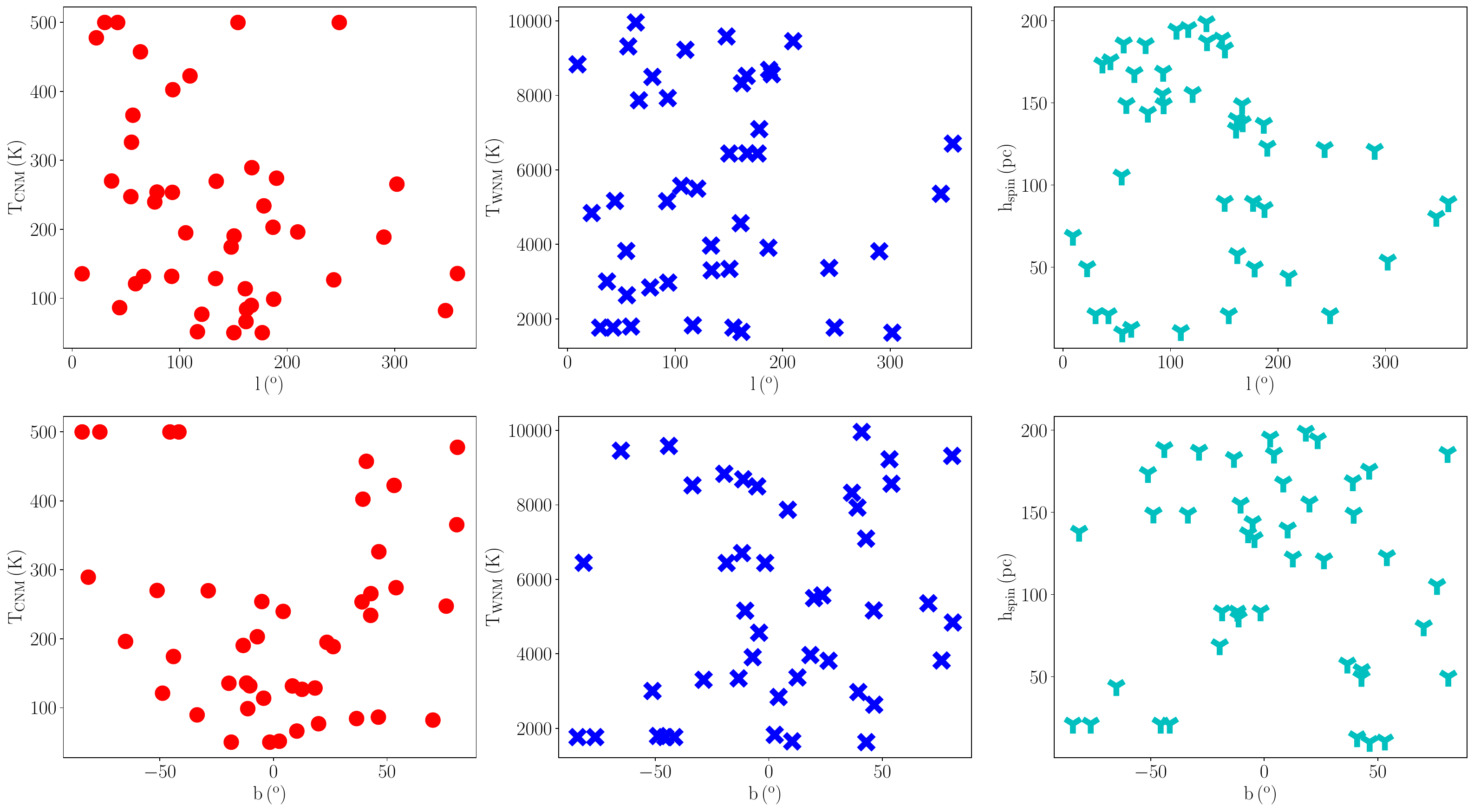}}
\end{center}
\caption{Variation of optimized spin temperatures and $h_{spin}$ as a function of $l$ and $b$. The top panels show optimized $T_{s,CNM}$ (left), $T_{s,WNM}$ (middle), and $h_{spin}$ (right) as a function of $l$. The bottom panels depict the same but as a function of $b$. As can be seen, optimized parameters show no dependence on $l$; however, the $T_{s,CNM}$ shows a dependence on $b$. See the text for more details.}
\label{fig:tau_params}
\end{figure*}

For our optimized model, we explore this variation in Fig.~\ref{fig:tau_params}, where we plot the optimized temperatures ($T_{s,CNM}$, $T_{s,WNM}$) and height ($h_{spin}$) as a function of $l$ and $b$. For a global two-temperature model to be true, the parameters should show minimal variation with $l$ and $b$. However, as shown in the figure, the optimized values span the full range. This indicates that a single two-temperature model with a thin cold disk with a fixed low spin temperature and a thick warm disk with a high spin temperature is not viable. The ISM has more complex structures with different $T_{s,CNM}$, and $T_{s,WNM}$ along different directions along with different filling factors. From the figure, we note that the optimized $T_{s,CNM}$ shows a moderate dependency on the galactic latitude, $b$. At low $b$ ($b \sim 0^o$), the optimized $T_{s,CNM}$ values get lower, and it increases gradually with increasing $b$ on both sides. This is consistent with the fact that at low $b$, the lines of sight cut a longer path close to the midplane.

\bibliographystyle{mnras}
\bibliography{bibliography}

\end{document}